\documentclass[twocolumn,preprintnumbers,superscriptaddress,endnote,nofootinbib,aps,prd,floatfix]{revtex4}

\usepackage[utf8]{inputenc}
\usepackage{footmisc,enumerate}
\usepackage{multirow}
\usepackage{subfigure}
\usepackage{amsmath,amsfonts,amssymb}
\usepackage{slashed}
\usepackage{graphicx}
\usepackage{placeins}
\usepackage{xspace}
\usepackage{hyperref}
\hypersetup{colorlinks=true, citecolor=blue, urlcolor=blue, linkcolor=blue}
\usepackage[normalem]{ulem}
\usepackage{sidecap}

\sidecaptionvpos{figure}{}

\makeatletter
\g@addto@macro\bfseries{\boldmath}
\makeatother

\begin{document}

\preprint{IPPP/19/71, MCnet}

\title{Higgs phenomenology as a probe of sterile neutrinos}

\begin{abstract}
Physics beyond the Standard Model can manifest itself as both 
new light states and heavy degrees of freedom. 
In this paper, we assume that the former comprise only a sterile neutrino, $N$.
Therefore, the most agnostic description of the new physics is given by 
an effective field theory built upon the Standard Model fields as well as $N$. 
We show that Higgs phenomenology provides a sensitive 
and potentially crucial tool to constrain effective gauge interactions 
of sterile neutrinos, not yet probed by current experiments. 
In parallel, this motivates a range of new Higgs decay channels 
with clean signatures as candidates for the next LHC runs, 
including $h\to \gamma+p_T^\text{miss}$ 
and $h\to \gamma\gamma+p_T^\text{miss}$.
\end{abstract}

\author{Jonathan M. Butterworth} \email{J.Butterworth@ucl.ac.uk}
\affiliation{Department of Physics \& Astronomy, University College London, 
London WC1E 6BT, United Kingdom\\[0.1cm]}
\author{Mikael Chala} \email{mikael.chala@durham.ac.uk}
\affiliation{Institute for Particle Physics Phenomenology, Durham University, Durham DH1 3LE, United Kingdom\\[0.1cm]}
\affiliation{CAFPE and Departamento de F\'isica Te\'orica y del Cosmos,
Universidad de Granada, E–18071 Granada, Spain\\[0.1cm]}
\author{Christoph Englert} \email{christoph.englert@glasgow.ac.uk}
\affiliation{SUPA, School of Physics \& Astronomy, University of Glasgow, Glasgow G12 8QQ, United Kingdom\\[0.1cm]}
\author{Michael Spannowsky} \email{michael.spannowsky@durham.ac.uk}
\affiliation{Institute for Particle Physics Phenomenology, Durham University, Durham DH1 3LE, United Kingdom\\[0.1cm]}
\author{Arsenii Titov} \email{arsenii.titov@durham.ac.uk}
\affiliation{Institute for Particle Physics Phenomenology, Durham University, Durham DH1 3LE, United Kingdom\\[0.1cm]}

\maketitle

\section{Introduction}
\label{sec:intro}
So far, no departure from Standard Model (SM) predictions has been established 
in collider experiments near or beyond the electroweak (EW) scale. This 
observation
suggests that any new physics beyond the SM (BSM) is either very weakly interacting,
or arises at a scale $\Lambda$ much larger than the electroweak scale; or 
both. While the scenario with only new heavy physics is successfully described 
using an effective field theory (EFT) framework, the so-called SMEFT 
\cite{Grzadkowski:2010es} (for a review see \cite{Brivio:2017vri}), in the 
latter case, where new physics manifests itself in the presence of very heavy 
resonances outside the kinematic reach of the LHC on the one hand and light very 
weakly coupled degrees of freedom on the other, to describe the resulting BSM 
phenomenology the EFT framework involves not only SM fields but also new degrees 
of freedom~--~which are likely singlets under the SM gauge 
group.%
\footnote{Although the possibility of
SM charged particles at the EW scale is
not fully ruled out yet~\cite{Egana-Ugrinovic:2018roi,Alcaide:2019kdr}, it is very unlikely.}

One popular scenario is the EFT of the SM extended with a sterile
neutrino $N$, also dubbed $\nu$SMEFT~\cite{delAguila:2008ir,Aparici:2009fh,Bhattacharya:2015vja,Liao:2016qyd}.%
\footnote{For the SM EFT extended with scalar singlets see \textit{e.g.} Refs.~\cite{Franceschini:2016gxv,Gripaios:2016xuo}.} 
Sterile neutrinos are present in many SM extensions, 
which aim to explain the origin of light neutrino masses. 
In particular, they are the main ingredient of  
the seesaw type I~\cite{Minkowski:1977sc,Yanagida:1979as,GellMann:1980vs,Glashow:1979nm,Mohapatra:1979ia} 
as well as the inverse~\cite{Mohapatra:1986aw,Mohapatra:1986bd,Bernabeu:1987gr} and linear~\cite{Akhmedov:1995ip,Akhmedov:1995vm,Malinsky:2005bi} seesaw mechanisms.
Although ``canonical'' (type~I) heavy neutrinos 
have masses close to the grand unification scale, 
mostly sterile neutrinos with much smaller masses can exist 
leading to various experimental signatures.
A variety of LHC studies have explored the phenomenology
of the $\nu$SMEFT for $N$ produced via contact
interactions~\cite{delAguila:2008ir,Duarte:2015iba,Duarte:2016caz,Alcaide:2019pnf}; 
in $W$ and top decays~\cite{BarShalom:2006bv,Cvetic:2018elt,Alcaide:2019pnf}
as well as via Higgs decays with $N$ decaying leptonically~\cite{Caputo:2017pit}; 
see also Ref.~\cite{Accomando:2016rpc}.

In this article, we focus on the production of one or two $N$
via the Higgs with each $N$ decaying
into a photon and missing energy. 
We show that current data are not
sensitive to the operators triggering these novel and clean Higgs signatures.
Moreover, we go beyond the aforementioned works on
this topic by performing much more realistic simulations of signals and background
and therefore of the LHC reach. We also comment on the sensitivity that can 
in principle be gained using data-driven approaches in these clean final states.

This article is organised as follows. We introduce the $\nu$SMEFT in
section~\ref{sec:framework} and single out those operators which are not yet constrained
by low-energy data. In section~\ref{sec:results} we discuss the potential of existing LHC searches and measurements to probe the aforementioned operators.
Likewise, in section~\ref{sec:monophoton} we propose dedicated searches in monophoton and di-photon Higgs decays.
We conclude in section~\ref{sec:conclusions}.

\section{Framework}
\label{sec:framework}
We consider the SM extended with one Majorana right-handed (RH) neutrino $N$ 
and assume that its mass is below the scale of new physics $\Lambda$. 
The renormalisable Lagrangian gets modified as follows:
\begin{equation}
\mathcal{L}^{d=4} = \mathcal{L}_\mathrm{SM} 
- \left[\lambda_i \overline{L_i} \tilde{H} N 
+ \frac{1}{2} m_N \overline{N^c} N + \mathrm{h.c.}\right],
\label{eq:Ldim4}
\end{equation}
%
where $\mathcal{L}_{SM}$ stands for the SM Lagrangian, $H$ represents the Higgs doublet, and $L$ is the doublet of left-handed (LH) leptons with $i = e,\mu,\tau$.
Following standard notation, we have defined $\tilde{H} = i \sigma_2 H^\ast$ and 
$N^c = C\overline{N}^T$ with $C$ being the charge conjugation matrix. Also,
$m_N$ is the Majorana mass of $N$.

Parameterising new physics effects in terms of higher-dimensional operators, 
at dimension five we have~\cite{Aparici:2009fh}
\begin{equation}
\mathcal{L}^{d=5} = \frac{\alpha_{LH}^{ij}}{\Lambda} \mathcal{O}_{LH}^{ij} 
+ \frac{\alpha_{NNH}}{\Lambda} \mathcal{O}_{NNH} 
+ \mathrm{h.c.}\,,
\end{equation}
%
where
\begin{align}
\mathcal{O}_{LH}^{ij} &= \overline{L_i^c}\tilde{H}^* \tilde{H}^\dagger L_j\,, 
\label{eq:OLH}\\
\mathcal{O}_{NNH} &= \overline{N^c}N H^\dagger H\,. 
\label{eq:ONNH}
\end{align} 
%
We note that for a single RH neutrino $N$ the operator
$\mathcal{O}_{NNB} = \overline{N^c} \sigma^{\mu\nu} N B_{\mu\nu}$ 
vanishes identically. 
At dimension six we consider the operators involving the Higgs doublet. 
The relevant Lagrangian reads
\begin{align}
\mathcal{L}^{d=6} &= \frac{\alpha_{HN}}{\Lambda^2} \mathcal{O}_{HN}
+ \bigg[\frac{\alpha_{LNH}^i}{\Lambda^2} \mathcal{O}_{LNH}^i 
+ \frac{\alpha_{HNe}^i}{\Lambda^2} \mathcal{O}_{HNe}^i \nonumber \\
&+ \frac{\alpha_{NB}^i}{\Lambda^2} \mathcal{O}_{NB}^i
+ \frac{\alpha_{NW}^i}{\Lambda^2} \mathcal{O}_{NW}^i
+ \mathrm{h.c.}\bigg]\,,
\end{align}
%
where~\cite{Liao:2016qyd}
\begin{align}
\mathcal{O}_{HN} &= \overline{N} \gamma^\mu N H^\dagger i \overleftrightarrow{D_{\mu}} H\,, 
\label{eq:OHN}\\
\mathcal{O}_{LNH}^i &= \overline{L_i} N \tilde{H} H^\dagger H\,, \\
\mathcal{O}_{HNe}^i &= \overline{N} \gamma^\mu e_{iR} \tilde{H}^\dagger i D_\mu H\,, \\
\mathcal{O}_{NB}^i &= \overline{L_i} \sigma^{\mu\nu} N \tilde{H} B_{\mu\nu}\,, \\
\mathcal{O}_{NW}^i &= \overline{L_i} \sigma^{\mu\nu} N \sigma_I \tilde{H} W^I_{\mu\nu}\,,
\label{eq:ONW}
\end{align}
%
and we have assumed the coefficients $\alpha$ to be real. 
In these equations, 
\begin{equation}
\sigma^{\mu\nu} = \frac{i}{2} \left[\gamma^\mu,\gamma^{\nu}\right]\,,
\end{equation}
%
$\sigma_I$ with $I=1,2,3$ are the Pauli matrices, and
\begin{equation}
H^\dagger \overleftrightarrow{D_\mu} H = H^\dagger D_\mu H - (D_\mu H)^\dagger H\,.
\end{equation}
%

For $m_N \lesssim 10$~keV, there are very stringent constraints 
on the new physics scale $\Lambda$ from cooling of red giant stars, 
implying $\Lambda \gtrsim 4 \times 10^6$~TeV~\cite{Aparici:2009fh}. 
In the range $10~\mathrm{keV} \lesssim m_N \lesssim 10~\mathrm{MeV}$, 
supernovae cooling produced by the transitions $\nu\gamma \to N$ 
provides the strongest bound, which depends on $m_N$ as
$\Lambda \gtrsim 4 \times 10^6 \times 
\sqrt{m_\nu/m_N}$~TeV~\cite{Aparici:2009fh}, 
where $m_\nu$ is the light neutrino mass. 
Taking $m_\nu \sim 0.01$~eV, we find $\Lambda \gtrsim 4\times10^3$~TeV 
(126~TeV) for $m_N = 10$~keV (10~MeV).

We are therefore interested in the regime in which $N$ is relatively light but
$0.01~\mathrm{GeV} \lesssim m_N \lesssim 10~\mathrm{GeV}$. The main decay 
channel is 
$N \to \nu \gamma$ induced by $\mathcal{O}_{NB}$ 
and $\mathcal{O}_{NW}$~\cite{Duarte:2015iba}. The corresponding decay rate  reads
\begin{equation}
\Gamma(N \to \nu_i \gamma) = \frac{m_N^3 v^2}{4\pi\Lambda^4} 
\left(\alpha_{NB}^i c_W + \alpha_{NW}^i s_W\right)^2.
\label{eq:Nvg}
\end{equation}
%
In principle, four-fermion operators are also present at dimension six, and 
can be expected to be more sizeable because they can arise at tree level. 
However, there are models in which four-fermion
operators are not generated at tree level; see appendix~\ref{sec:app}.  Moreover, 
even if present these operators
do not interfere with Higgs processes, which are the ones we are more interested 
in. Likewise, $N\to\nu\gamma$ is the dominant decay 
irrespectively of the value of four-fermion interactions in the range of mass 
under consideration.

In this range of $m_N$ and assuming a standard cosmological history,
the contribution of the new neutrino to $N_\text{eff}$ does not	saturate the 
current Planck limit
$\Delta N_\text{eff}\lesssim 0.3$~\cite{Aghanim:2018eyx}.
(In alternative cosmologies, for	example	if the reheating temperature is	
close to $\sim 10$~MeV,
$N$ does not achieve thermal equilibrium with the SM fields at any time, 
and CMB constraints can be even more easily avoided~\cite{Abazajian:2019oqj}.)

We will make a number of assumptions on the coefficients of 
the considered operators 
to avoid constraints from low-energy data.
First of all, we neglect the Yukawa couplings $\lambda_i$ in 
Eq.~\eqref{eq:Ldim4}, 
which after electroweak symmetry breaking (EWSB)
generate mixing of $N$ with the SM neutrinos. 
Through the operators $\mathcal{O}_{NB}^i$ and $\mathcal{O}_{NW}^i$ 
this mixing would induce magnetic moments for the SM neutrinos, 
which are strongly constrained by reactor, accelerator 
and solar neutrino data~\cite{Canas:2015yoa,Miranda:2019wdy}.
This is entirely due to the missing $t$-channel mass suppression of the photon 
exchange.%
\footnote{On the other hand, 
massive mediators are largely unconstrained. 
Accelerator experiments ({\textit{e.g.}}~\cite{Auerbach:2001wg})
are relevant in a mass region of around less than $50~\text{MeV}$,
while masses in the keV range are subject to tight reactor data constraints, {\textit{e.g.}}~\cite{Beda:2012zz}. 
These are mass scales which are far below the typical 
hadron collider momentum transfers of ${\cal{O}}(100~\text{GeV})$ once trigger and selection criteria are included, which means
that our results are insensitive to the concrete mass choice of neutrinos with masses $\lesssim 1~\text{GeV}$. In this context,
the low-energy measurements are only relevant when we make a concrete choice of small masses that do not
impact our LHC analyses for the range that we consider. We will therefore not 
include the low-energy constraints explicitly in this work.}

Naively, even if $\lambda_i = 0$, the mixing would be induced 
after EWSB by the dimension-six operators $\mathcal{O}_{LNH}^i$. 
However, without loss of generality, 
we can redefine the couplings $\lambda_i$ in Eq.~\eqref{eq:Ldim4} 
as $\lambda_i \rightarrow \lambda_i + \alpha_{LNH}^i v^2/(2\Lambda^2)$ 
from the beginning, $\langle H^0 \rangle = v/\sqrt{2}$ being the Higgs 
vacuum expectation value (VEV). Such parameterisation of the 
Yukawa couplings ensures that setting $\lambda_i = 0$ 
leads to no mixing, and we assume this in what follows.%
\footnote{If the condition $\lambda_i \approx 0$ holds at a scale $\Lambda \gg v$, then the different RGE
running of the Yukawa and the $\mathcal{O}_{LNH}$ operators might induce a SM neutrino dipole
moment, $\mathcal{O}_{\nu A} = (\overline{\nu_{iL}}\sigma^{\mu\nu}\nu_{jL}^c) A_{\mu\nu}$. The size of
this operator can be estimated to be
$$\alpha_{\nu A}\sim \frac{\alpha_{LNH}\alpha_{NA}g^2}{(4\pi)^2}\frac{v^3}{\Lambda^4} \log{\frac{\Lambda}{v}}\,.$$ 
On the other hand, the latter is experimentally bounded to be $\alpha_{\nu A}\lesssim 2\times 10^{-14}$ TeV$^{-1}$~\cite{Canas:2015yoa,Bell:2006wi}.
Therefore, $\lambda_i$ must vanish close to the EW scale (note that, below the EW scale, the
dipole moment does not renormalise~\cite{Jenkins:2017dyc}). Otherwise, $\alpha_{LNH} \lesssim 10^{-10}$ (for $\alpha_{NA}\sim 1$ and $\Lambda\sim 1$ TeV), making searches for processes triggered by $\mathcal{O}_{LNH}$ irrelevant. (Searches triggered by $\mathcal{O}_{NNH}$ would still 
be relevant in this regime, though.) This strong bound on $\alpha_{LNH}$ is however no longer valid if $N$ couples to only one lepton family (because for Majorana neutrinos the dipole operator must be antisymmetric); this happens \textit{e.g.} if lepton number is approximately conserved in each family, as it occurs within the SM. (In such case, though, $\alpha_{NNH}$ would be small.)}
Then, the Higgs-neutrino interaction reads
\begin{equation}
\frac{\alpha_{LNH}^i}{\sqrt{2}} \frac{v^2}{\Lambda^2} \overline{\nu_{iL}}Nh 
+ \mathrm{h.c.}
\end{equation}
%
Since the term of $O_{LNH}^i$ proportional to $v^3$ cancels out, 
these operators contribute neither to the $W^\pm \to \ell^\pm N$ 
nor the $Z \to \nu N$ decay widths.

Upon EWSB, the operator $\mathcal{O}_{NNH}$ contributes to 
the Majorana mass of $N$. Similarly to the discussion above, 
we can redefine $m_N$ in Eq.~\eqref{eq:Ldim4} as 
$m_N \rightarrow m_N + \alpha_{NNH} v^2/\Lambda$. 
In this way, $m_N$ is the physical mass of $N$. 
The $hNN$ interaction arising from this operator has the form
\begin{equation}
\alpha_{NNH} \frac{v}{\Lambda} \overline{N^c}Nh + \mathrm{h.c.}
\end{equation}
%

We attribute the smallness of $\nu_i$ masses 
to the extremely small values of the coefficients $\alpha_{LH}^{ij}$ 
of the Weinberg operator given in Eq.~\eqref{eq:OLH}. 
Therefore, we do not consider this operator in what follows. 

In light of the previous discussion, we focus on the operators 
given in Eqs.~\eqref{eq:ONNH} and \eqref{eq:OHN}--\eqref{eq:ONW}.
The operator $\mathcal{O}_{HN}$ leads to the decay $Z \to NN$ with 
the following width (we neglect $m_N$ in analytical computations below):
\begin{equation}
\Gamma(Z \to NN) = \frac{m_Z^3 v^2}{24\pi\Lambda^4} \alpha_{HN}^2\,. 
\end{equation}
%
The operators $\mathcal{O}_{NB}^i$ and $\mathcal{O}_{NW}^i$ 
give rise to
\begin{equation}
\Gamma(Z \to \nu N) = \frac{m_Z^3 v^2}{12\pi\Lambda^4}
\sum_i \left(\alpha_{NW}^i c_W - \alpha_{NB}^i s_W \right)^2\,.
\label{eq:ZnuN}
\end{equation}
%
The operators $\mathcal{O}_{HNe}^i$ and $\mathcal{O}_{NW}^i$ 
contribute to the $W$ decay width: 
\begin{equation}
\Gamma(W^+\to \ell_i^+ N) = \frac{m_W^3 v^2}{48\pi\Lambda^4} 
\left[(\alpha_{HNe}^i)^2 + 4 (\alpha_{NW}^i)^2 \right].
\label{eq:WlN}
\end{equation}
%

It proves convenient for our further discussion 
to define the following operators: 
\begin{align}
\mathcal{O}_{NA}^i &=\phantom{-} c_W \mathcal{O}_{NB}^i + s_W \mathcal{O}_{NW}^i\,,\\
\mathcal{O}_{NZ}^i &= -s_W \mathcal{O}_{NB}^i + c_W \mathcal{O}_{NW}^i\,,
\end{align}
%
where $c_W \equiv \cos\theta_W$, $s_W \equiv \sin\theta_W$, 
with $\theta_W$ being the Weinberg angle. 
Then we can rewrite 
\begin{equation}
\alpha_{NB}^i \mathcal{O}_{NB}^i + \alpha_{NW}^i \mathcal{O}_{NW}^i 
= \alpha_{NA}^i \mathcal{O}_{NA}^i + \alpha_{NZ}^i \mathcal{O}_{NZ}^i\,,
\end{equation}
%
where
\begin{align}
\alpha_{NA}^i &\equiv \alpha_{NB}^i c_W + \alpha_{NW}^i s_W\,, 
\label{eq:alphaNA}\\
\alpha_{NZ}^i &\equiv \alpha_{NW}^i c_W - \alpha_{NB}^i s_W\,.
\label{eq:alphaNZ}
\end{align}
%

Given these equations, we focus on a regime with $\alpha_{HN} = 0$. 
This coefficient is extremely constrained by 
the measurement
$\Gamma (Z \to \nu\nu\gamma\gamma)/\Gamma_Z^\mathrm{total}
< 3.1\times 10^{-6}$~\cite{Tanabashi:2018oca}. (In our set-up,
$\Gamma(Z \to \nu\nu\gamma\gamma) 
= \Gamma(Z \to NN) \mathcal{B}^2(N \to \nu\gamma)$ 
and $\mathcal{B}(N \to \nu\gamma) \approx 1$ 
for the values of $m_N$ of interest.)

Secondly, we set $\alpha_{NZ}^i = 0$, which implies 
$\Gamma(Z \to \nu N) = 0$ (see Eq.~\eqref{eq:ZnuN}). 
In this way we avoid the strong constraints on $Z \to \gamma + p_T^\text{miss}$~\cite{Adriani:1992iu,Akers:1994vh,Abreu:1996vd,Acciarri:1997im}.
Finally, we also assume that $\alpha_{HNe}^i = 0$, such that 
we completely avoid bounds from measurements of the $W$ width. 
Moreover, the operators $\mathcal{O}_{HNe}^i$ do not contribute to the processes we analyse in this work, so 
we can set $\alpha_{HNe}^i=0$ without loss of generality.

Under these assumptions we can express from Eq.~\eqref{eq:alphaNZ}
$\alpha_{NW}^i = \alpha_{NB}^i t_W$, where $t_W \equiv s_W/c_W$, 
and rewrite Eq.~\eqref{eq:WlN} in terms of $\alpha_{NA}^i$ as
\begin{equation}
\Gamma(W^+ \to \ell_i^+ N) = \frac{m_W^3 v^2}{12\pi\Lambda^4} 
s_W^2 (\alpha_{NA}^i)^2.
\end{equation}
%
Thus, we are left with only $\alpha_{NA}^i$, $\alpha_{LNH}^i$ and $\alpha_{NNH}$. 
The value of $\alpha_{NA}^i/\Lambda^2$ is constrained by 
the measurement of 
$\Gamma_W^\mathrm{total} = 2.085 \pm 0.042$~GeV 
to be $(\alpha_{NA}^i/\Lambda^2) \lesssim 4\pi~\mathrm{TeV}^{-2}$, 
so we can vary it in $[0.001, 4\pi]$ for $\Lambda = 1$~TeV. 
We have assumed that all three $\alpha_{NA}^i$ are of the same order. 
The lower bound is set by the requirement that $N$ decays 
promptly enough (within 4 cm), see Eq.~\eqref{eq:Nvg}. 
Other low energy constraints on $\alpha_{NA}$ are very weak; 
see \textit{e.g.} Ref.~\cite{Magill:2018jla}.%
\footnote{For $m_N \lesssim 100$ MeV, $\alpha_{NA}$ might be expected to trigger the pion decay $\pi^\pm\to \ell^\pm N$, and therefore be severely constrained. However, the corresponding amplitude scales as~\cite{Alcaide:2019pnf} $$\mathcal{M}\sim \langle 0|V^\mu|\pi^\pm \rangle (\gamma_\mu\slashed{p}-p_\mu)\,;$$ which vanishes because $\langle 0|V^\mu|\pi^\pm \rangle = f_\pi p^\mu$, with $p$ being the pion four-momentum.}
For $\Lambda = 1$~TeV, the coefficients 
$\alpha_{LNH}^i$ and $\alpha_{NNH}$ can run in the range 
$[0,0.5]$ and $[0, 0.05]$, respectively. 
The stringent upper bounds on these coefficients 
follow from the requirement that the partial decay widths of the Higgs boson
in Eqs.~\eqref{eq:HvN} and \eqref{eq:HNN} 
discussed later do not 
exceed 
the total Higgs width in the SM, $\Gamma_H^\mathrm{total} \approx 4$~MeV.

\section{Existing searches}
\label{sec:results}
The most important processes at the LHC triggered by those operators which are very weakly constrained by low-energy data, namely 
$\mathcal{O}_{NA}^i$, $\mathcal{O}_{LNH}^i$ and $\mathcal{O}_{NNH}$ are:
\begin{itemize}
\item $pp \to \gamma^\ast \to \nu \nu \gamma$ (through $\mathcal{O}_{NA}^i$), 
meaning $pp \to \gamma^\ast \to \nu N$ plus subsequent decay $N \to \nu \gamma$;
\item $pp \to W^\pm \to \ell^\pm \nu \gamma$ (through $\mathcal{O}_{NA}^i$), 
meaning $pp \to W^\pm \to \ell^\pm N$ plus subsequent decay $N \to \nu \gamma$;
\item $pp \to h \to \nu \nu \gamma$ (through $\mathcal{O}_{LNH}^i$), meaning 
$pp \to h \to \nu N$ plus subsequent decay $N \to \nu \gamma$;
\item $pp \to h \to \nu \nu \gamma \gamma$ (through $\mathcal{O}_{NA}^i$), 
meaning $pp \to h \to \nu N \gamma$ plus subsequent decay $N \to \nu \gamma$;
\item $pp \to h \to \nu \nu \gamma \gamma$ (through $\mathcal{O}_{NNH}$),
meaning $pp \to h \to N N$ plus decay of each $N \to \nu \gamma$.
\end{itemize}
%
(Note that the first and third processes do not interfere 
because the Higgs production is mostly initiated by gluons, 
while Drell-Yan production is initiated by quarks.) 
Let us first focus on the neutral current Drell-Yan process.

In the limit of vanishing masses, the differential cross section for 
$q\overline{q} \to \nu_i N$ is found to be 
\begin{equation}
\frac{\mathrm{d}\sigma}{\mathrm{d}t}(q\overline{q} \to \gamma^\ast \to \nu_i N) = 
- \frac{2\alpha Q^2 v^2}{3\Lambda^4s^3} \left(\alpha_{NA}^i\right)^2 t\left(s+t\right)\,,
\end{equation}
%
where $\alpha = e^2/(4\pi)$ is the fine-structure constant,
and $Q$ is the electric charge of the quark $q$. 
The integrated cross section 
\begin{equation}
\sigma(q\overline{q} \to \gamma^\ast \to \nu_i N) = 
\frac{\alpha Q^2 v^2}{9\Lambda^4} \left(\alpha_{NA}^i\right)^2
\end{equation}
%
is independent of $s$. 
For the process of interest we have
$\sigma(q\overline{q} \to \gamma^\ast \to \nu \nu \gamma) = 
\sigma(q\overline{q} \to \gamma^\ast \to \nu N) 
\mathcal{B}(N \to \nu \gamma)$, where 
$\mathcal{B}(N \to \nu \gamma) \approx 1$ 
for the considered range of $m_N$. 

This process can be constrained at the LHC 
in searches for events with one photon and missing energy. 
To the best of our knowledge, the most up-to-date search in this
respect is the CMS analysis of Ref.~\cite{Sirunyan:2018dsf}.
Most importantly, this analysis requires exactly one photon with
$p_T^\gamma > 175$ GeV and $|\eta_\gamma|<1.44$ as well as missing energy $p_T^\text{miss} > 170$ GeV. The ratio
$p_T^\gamma/p_T^\text{miss}$ is required to be smaller than 1.4
in order to reduce the background from $\gamma+\text{jets}$. 
With the same aim, events are rejected if the minimum opening angle 
between $p_T^\text{miss}$ and the transverse momentum of the four
hardest jets is less than 0.5. (Only jets with $p_T^j>30$ GeV
and $|\eta_j|<5$ are considered in this cut.) Likewise, $\Delta\phi(p_T^\gamma,p_T^\text{miss}) > 0.5$.
Finally, events are also rejected if they contain any electron
or muon with $p_T> 10$ GeV within $\Delta R > 0.5$ from the photon.

The analysis considers two signal regions depending on whether
$|\sin{\phi}|$ is smaller or larger than $\sin{(0.5)}$ or not, which are further split
in six $p_T^\gamma$ bins in the range  
$[175,1000]$~GeV; see Tab.~1
in the experimental report~\cite{Sirunyan:2018dsf}.

We recast this search using dedicated routines based on \texttt{ROOT v5}~\cite{Brun:1997pa,Antcheva:2009zz},
\texttt{HepMC v2}~\cite{Dobbs:2001ck} and \texttt{FastJet v3}~\cite{Cacciari:2011ma}.
Jets are
built using the anti-k$_t$ algorithm with $R=0.4$, and defined by
$p_T^j > 10$ GeV. For photons and leptons we require $p_T > 10$ GeV. 
These objects are experimentally under very good control~\cite{Hoya:2628323}.

We find that the most constraining signal region is that with
$|\sin{\phi}| < \sin{(0.5)}$ and $p_T^\gamma\in[300, 400]$ GeV.
The experimental collaboration reports the observation of 44 events,
while $46.6\pm 4.0$ are predicted in the SM. Using the CL$_s$ method~\cite{Read:2000ru,Read:2002hq},
including the uncertainties in the estimation of the SM background,
we obtain that the maximum number of signal events in this bin is 16.

We estimate that the efficiency for selecting signal events in this bin
in Drell-Yan processes triggered by $\mathcal{O}_{NA}$ is $\sim 0.057$. 
Using the LO production cross section before cuts, 
we obtain that $\alpha_{NA} > 0.88$ is excluded already at the 95\% CL, 
assuming that $N$ couples to only one lepton family.

\begin{figure}[t]
\centering
\includegraphics[width=0.45\textwidth]{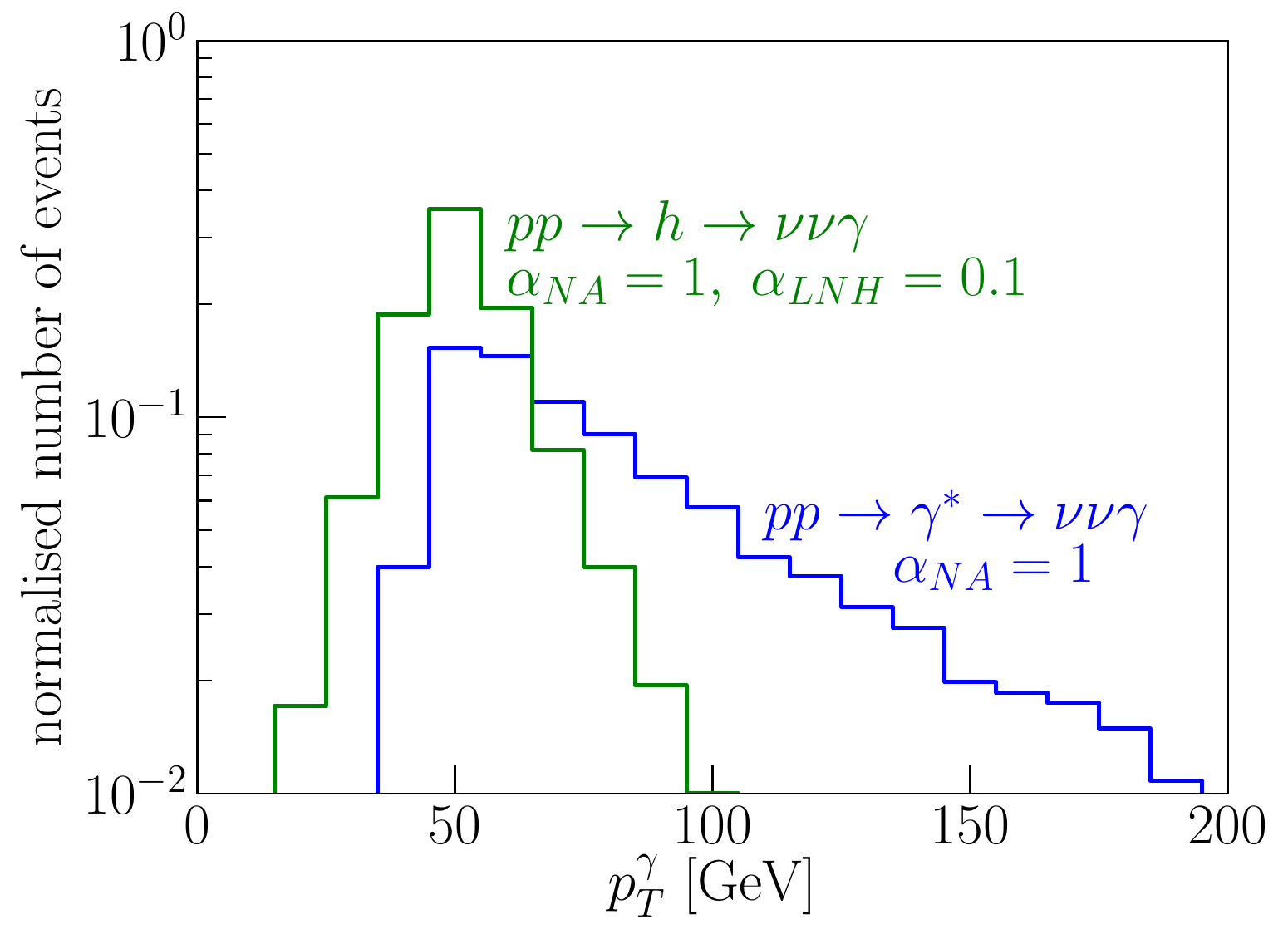}
\caption{\it The $p_T^\gamma$ distributions for 
$pp \to \gamma^\ast \to \nu \nu \gamma$ and 
$pp \to h \to \nu \nu \gamma$. 
We have generated events setting $\alpha_{NA} = 1$ for both processes, 
and in addition $\alpha_{LNH} = 0.1$ for the second process. 
All the other coefficients have been set to zero.}
\label{fig:monophoton_pT}
\end{figure}

Interestingly, when running the simulated analysis over events of the type 
$pp\rightarrow h\rightarrow \nu\nu\gamma$, we find that none of the bins 
in this search constrains the operator $\mathcal{O}_{LNH}$. 
This can be easily understood because the distribution 
of the transverse momentum of the photon falls much faster in this process; 
see Fig.~\ref{fig:monophoton_pT}.
%
\begin{SCfigure*}[20][t]
\centering
\includegraphics[width=0.7\textwidth]{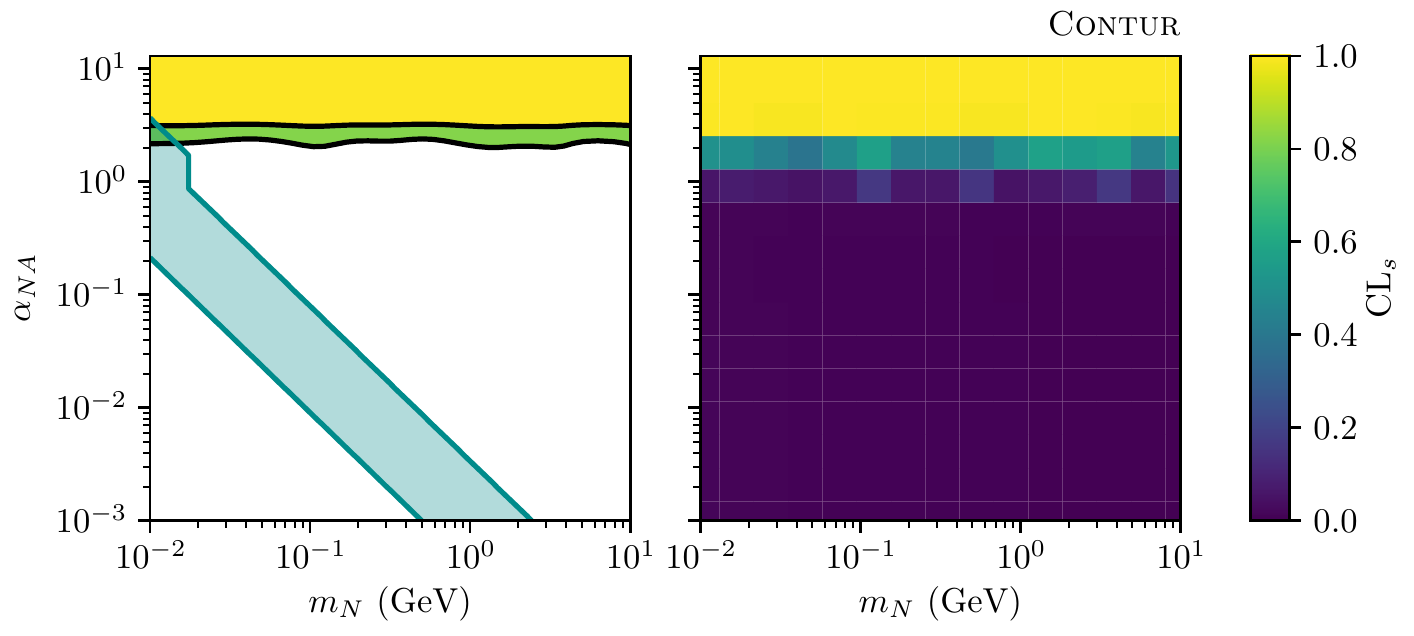}
\caption{\it\textsc{Contur} exclusion in the $\alpha_{NA}, m_N$ plane, 
for $\Lambda = 1$~TeV. The left-hand inset shows the 2- and 1-$\sigma$
exclusion contours based on the heatmap on the right. 
Within the diagonal cyan region the heavy neutrino would decay 
within the detector volume; above, it is effectively prompt 
and below, it is effectively stable.}
\label{fig:contur}
\end{SCfigure*}
%

On another front, differential cross section measurements have been made 
at the LHC for a wide range of potentially relevant final states. 
These measurements are generally made in well-defined fiducial kinematic 
regions, giving them a high degree of model-independence, 
and corrected for detector effects 
(such as resolution and reconstruction efficiency) 
within these regions, meaning they can be readily compared 
to generated signal events. 
We have used the \textsc{Contur}~\cite{Butterworth:2016sqg} tool 
to study whether our model would have had a visible impact on 
any of these measurements, all of which are currently consistent with the SM. 
To do this, we use the 
\texttt{Herwig 7.1.5}~\cite{Bellm:2015jjp,Bahr:2008pv} event generator 
to read the \texttt{UFO}~\cite{Degrande:2011ua} files of our model and produce simulated collision events. 
All processes with any BSM content in the final state, or 
on-shell intermediate states, are generated at LO tree level. For each parameter 
point, one million events are generated, 
implying an integrated luminosity at least equivalent to that of the data, and typically much higher. 
These events are then passed to {\sc{Rivet}}~\cite{Buckley:2010ar}, which contains 
implementations of a large number of the relevant LHC analyses as well as the measurement data derived from 
{\sc{HepData}}~\cite{Maguire:2017ypu}. The effect of injecting signal events on top of the data for all these measurements 
is then evaluated by \textsc{Contur}, and the most sensitive distributions for any given model point are identified and
used to derive a potential exclusion, using a $\chi^2$ test.

Scanning the range $10^{-3} < \alpha_{NA} < 4\pi$ with the other couplings set to zero, we observe that for 
$\alpha_{NA} \gtrsim 3$, heavy neutrino production via Drell-Yan is significant, $q^\prime\bar{q} \rightarrow N e^\pm$
and $q\bar{q} \rightarrow N \nu_e$.
For example, at $\alpha_{NA} = 3.4$ and $m_N = 0.2$~GeV, the inclusive cross 
section for these processes combined is
40~pb, dominated by the channels involving $e^\pm$. As they contain an electron and missing transverse energy, these
events can easily populate the fiducial phase space of measurements aimed at $W$-bosons decaying to electrons and 
neutrinos~\cite{Aad:2014qxa}, 
with the photon from the $N$ decay
also meaning that they can impact upon $W+\gamma$ measurements~\cite{Aad:2013izg}. They, and the neutral current Drell-Yan processes, 
can also contribute to inclusive photon and photon-plus-jet measurements~\cite{Aad:2016xcr,Aaboud:2017kff}, where no veto is made on
the rest of the event, and photon-plus-missing-energy measurements~\cite{Aad:2016sau}.
The resultant exclusion is shown in Fig.~\ref{fig:contur}.
More recent measurements, and those from CMS in
these channels, are not yet included in {\sc{Rivet}} and so are not used. 

We note there is no dependence on $m_N$ over the range considered.
Setting $\alpha_{NA} = 1$, $\alpha_{NNH} =0$ and scanning 
$0 < \alpha_{LNH} < 0.5$ for the same range of $m_N$, 
some events do enter the fiducial region of the same measurements, 
but there is no impact at the $1\,\sigma$ or above level. 
The same is true for scanning  $0 < \alpha_{NNH} < 0.05$ 
with $\alpha_{NA} = 1$, $\alpha_{LNH} =0$. 
The fact that some events do populate the acceptance of these measurements 
means there may be sensitivity as the measurement precision is increased with
higher integrated luminosity.

These existing limits leave a large part 
of well-motivated parameter space uncovered. 
To probe this region, new analyses, not yet conceived, are 
required.%
\footnote{The very recent search for 
$Zh$ with $h\to\gamma+p_T^\text{miss}$ by CMS~\cite{Sirunyan:2019xst} has still very little sensitivity.}
We discuss such analyses in the next section.

\section{Higgs searches in the mono and di-photon + missing energy channels}
\label{sec:monophoton}
In this section we investigate the sensitivity reach of the LHC 
to the mono and di-photon + missing energy channels 
through novel Higgs decays. 

There are different strategies
to constrain new physics signals at colliders. On the one hand, if a good understanding of the background and 
the signal can be achieved this can be used to inform an experimental search in cut-and-count or
more sophisticated multivariate analyses, in line with the previous section. 
This approach is the major driving force behind searches in complicated
multi-scale final states.
As also discussed in the previous section,
precision differential measurements of relatively simple final states can also contribute.
\begin{figure*}[t]
\centering
\subfigure[~]{\includegraphics[width=0.45\textwidth]{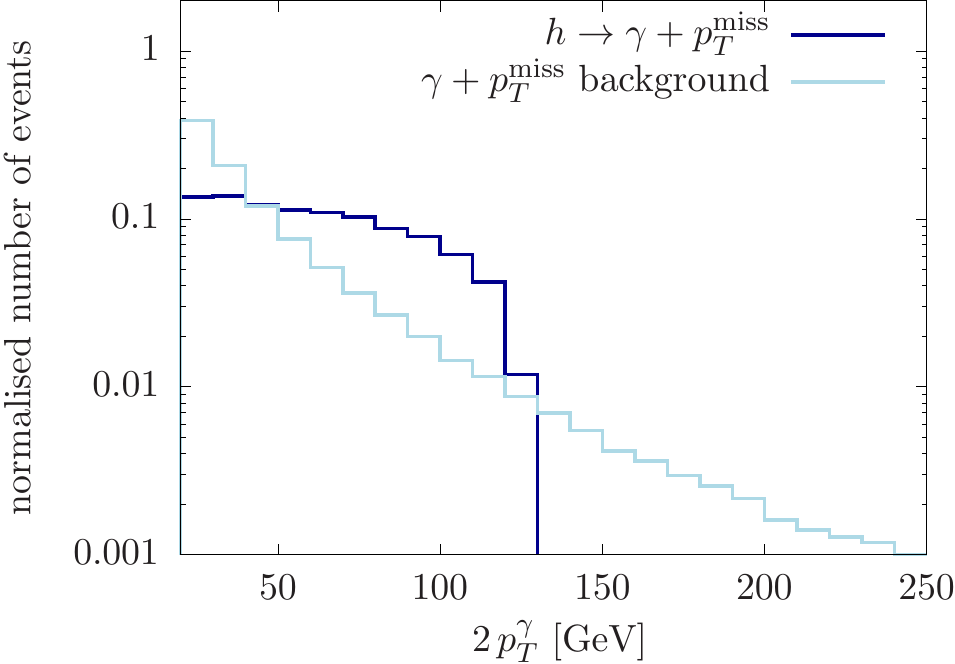}}
\hspace{1cm}
\subfigure[~]{\includegraphics[width=0.45\textwidth]{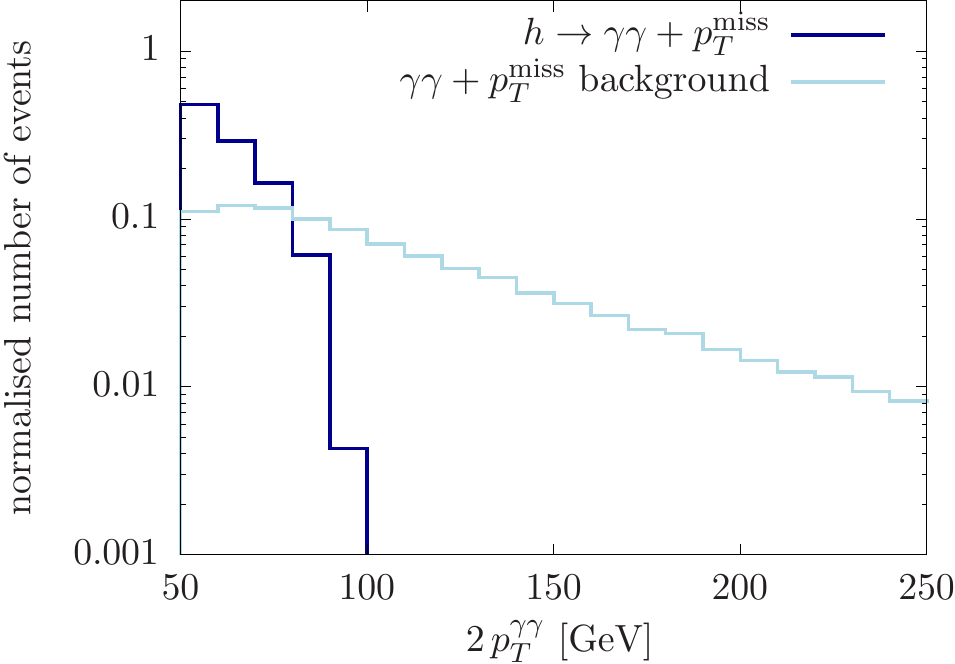}}
\caption{\it (a) Shape of the transverse momentum distributions of the photon (inflated by a factor of two, hence describing the transverse mass) in $h\to \gamma+p_T^{\text{miss}}$. 
The background is generated using {\sc{MadGraph}} and is used as placeholder of the data-driven analysis approach discussed in the text. (b) Same as (a) but for  the di-photon transverse momentum distributions for $h\to \gamma\gamma+p_T^{\text{miss}}$.}
\label{fig:photon} 
\end{figure*}
%

On the other hand, if a signal process is clustered in a particular phase
space region (\textit{e.g.} for resonance searches) we can use sidebands to constrain the background with
a minimum of theoretical input using data-driven approaches. Such a strategy comes into
its own when the final state objects are experimentally under good statistical and systematic control, which is
the case for photons already at low transverse momenta~\cite{Hoya:2628323}. If the shape of a background can be
accurately described using fitting techniques across a large range of a kinematical observable, such a strategy
can be used to detect small signals on top of large backgrounds even when the latter are theoretically not well-understood. 
A prime example of this strategy is arguably the Higgs discovery 
via its decay to $\gamma\gamma$ with relatively small 
signal-to-noise ratio~\cite{Aad:2012tfa,Chatrchyan:2012xdj}. 

Our search, although involving missing energy with different systematic properties, shares many similarities with the $h \to \gamma\gamma$
search: the background is large and the expected branching ratio 
$h\to \gamma(+\gamma)+p_T^{\text{miss}}$ is small, \textit{i.e.} the new
physics signal is likely to be dwarfed by the expected theoretical uncertainties associated with mono and di-photon production in
a complicated hadronic environment. However, the signal normalisation
is accurately known~(see \textit{e.g.} Refs.~\cite{Dittmaier:2011ti,Dittmaier:2012vm,Heinemeyer:2013tqa,deFlorian:2016spz}) and its relevant
final state kinematics is entirely determined by the Higgs mass which can be accurately extracted from subsidiary measurements; see for example the ATLAS and CMS combination of Ref.~\cite{Aad:2015zhl}). This implies a distinct shape of the new physics signal, \textit{i.e.} there is Jacobian peak in the transverse
momentum distribution of the photon or photon pair, depending on which final state we are interested in as shown in Fig.~\ref{fig:photon}. Note that this way the resonance cross section extraction is also not impacted by BSM contributions to the continuum as indicated in Fig.~\ref{fig:monophoton_pT}: the background shape might change but the resonance will still have a distinct shape, which can be extracted from the continuum for large enough data sets.

In the following we take inspiration from the $h\to \gamma\gamma$ search and estimate the sensitivity at the high-luminosity LHC by performing a template fit on the signal and background distributions using {\tt{RooStats}}~\cite{Moneta:2010pm} (background shape estimates taken from Monte Carlo), leaving their normalisations as free parameters. (Using the same approach, we are able to reproduce \textit{e.g.} the expected ATLAS $p$-value of the 8 TeV $h\to \gamma\gamma$ search of Ref.~\cite{Aad:2012tfa} within 10\%. This highlights that such an approach is highly feasible when all experimental aspects are under good systematic control, which we assume here implicitly, but not unrealistically.) The 95\% CL constraint on the signal modifier when agreement with the background-only hypothesis is given can then be understood as a direct constraint on the respective branching
ratios when using the signal normalisation of $pp \to h$ from~Ref.~\cite{deFlorian:2016spz}. 
The expected background cross sections for the inclusive selection criteria that underpin Fig.~\ref{fig:photon} are $\sigma(\gamma+p_T^{\text{miss}})\simeq 14~\text{pb}$ and $\sigma(\gamma\gamma+p_T^{\text{miss}})\simeq 10~\text{fb}$. This way we obtain 
\begin{align}
\mathcal{B}(h\to \gamma + p_T^{\text{miss}}) & = 1.2 \times 10^{-4}\,,
\label{eq:Bmonophoton} \\
\mathcal{B}(h\to \gamma \gamma + p_T^{\text{miss}}) & = 4.2 \times 10^{-5}\,,
\label{eq:Bdiphoton}
\end{align}
%
at $3$ ab$^{-1}$
using signal and background templates generated with {\sc{MadGraph}}~\cite{Alwall:2014hca} as shown in Fig.~\ref{fig:photon}. In line with the previous section we require a minimum $p_T$ of the photon of 10 GeV for our mock $\gamma+p_T^{\text{miss}}$ data and the expected SM $h\to Z\gamma$ contribution is subtracted from these numbers. 
Note that owing to the much smaller expected background of the $\gamma\gamma+p_T^{\text{miss}}$ analysis, the signal is naively easier to isolate, but it is considerably more washed out due to missing energy systematics. These issues fall into the area of experimental expertise, hence we limit ourselves to the sensitivity estimated along the lines above, but we choose harder photons $p_T\geq 15~\text{GeV}$ with separation in the pseudo-rapidity--azimuthal angle plane of at least 0.4 as well as a minimum missing transverse energy of 25~GeV for the di-photon analysis to partially take into account the more complicated nature of this process and report results separately.

The bounds in Eqs.~\eqref{eq:Bmonophoton} and \eqref{eq:Bdiphoton} can be 
translated to constraints on $\alpha_{LNH}$, $\alpha_{NNH}$ and $\alpha_{NA}$ by 
means of 
\begin{align}
\Gamma(h \to \nu N) &= \frac{ m_h v^4}{16\pi\Lambda^4} \sum_i (\alpha_{LNH}^i)^2\,, 
\label{eq:HvN} \\
\Gamma(h \to NN) &= {m_h v^2 \over 4\pi \Lambda^2} \alpha_{NNH}^2\,,
\label{eq:HNN} \\
\Gamma(h\rightarrow \gamma \nu N) &= \frac{ 
m_h^5}{768\pi^3\Lambda^4} 
\sum_i \left(\alpha_{NA}^i \right)^2\,.
\label{eq:NA}
\end{align}
%
(See Ref.~\cite{Shrock:1982kd} for former partial computations.)
Each of the three Wilson coefficients can be bounded independently 
by setting the remaining two to zero. 
(Note that any other choice would lead to a more stringent constraint.) 
The sensitivity at the high-luminosity LHC can be read in Tab.~\ref{tab:limits}. 
\begin{table}[t]
\begin{tabular}{|c|c|c|c|}
\hline
\multirow{2}{*}{Operator} & $\alpha_{\text{max}}$ & $\Lambda_{\text{min}}$~[TeV]  & \multirow{2}{*}{Channel}\\
& for $\Lambda=1$ TeV & for $\alpha=1$ & \\ \hline 
$\mathcal{O}_{LNH}$ & $4.2\times 10^{-3}$ & $15$ & $h\to\gamma+p_T^{\text{miss}}$\\
$\mathcal{O}_{NNH}$ & $5.3\times 10^{-4}$ & $1900$ & $h\to\gamma\gamma+p_T^{\text{miss}}$\\
$\mathcal{O}_{NA}$ & $0.21$ & $2.2$ & $h\to\gamma\gamma+p_T^{\text{miss}}$\\
\hline
\end{tabular}
\caption{\it Maximum (minimum) value of 
$\alpha~(\Lambda)$ for $\Lambda = 1~\text{TeV}~(\alpha = 1)$ 
allowed by the proposed searches quoted in the last column. 
We have assumed lepton flavour universality in couplings to $N$.}
\label{tab:limits}
\end{table}
%
We remind the reader that all these prospects apply only if 
$\alpha_{NA}/\Lambda^2 \gtrsim 0.001- 0.1$~TeV$^{-2}$, 
depending on $m_N$; see Fig.~\ref{fig:contur}.

\section{Conclusions}
\label{sec:conclusions}
In summary, we have studied the phenomenology of the SMEFT extended with a light RH neutrino $N$ in the regime in which the latter decays almost exclusively into a photon and a neutrino.

Using low-energy and LHC data such as measurements of the $W$, $Z$ and Higgs bosons; bounds on neutrino dipole moments, measurements of SM differential distributions at the LHC (as implemented in \textsc{Contur}~\cite{Butterworth:2016sqg}), as well as searches for single photons with missing energy~\cite{Sirunyan:2018dsf}; we have singled out those directions not yet constrained. They include mostly operators triggering new Higgs decays, namely $h\to \gamma +p_T^{\text{miss}}$ and $h\to\gamma\gamma+p_T^{\text{miss}}$.

We have subsequently provided new search strategies to be performed 
at the LHC sensitive to the aforementioned processes. 
For order one couplings, we have shown that, with 3 ab$^{-1}$ of data, 
these analyses can potentially unravel new physics at scales 
$\Lambda \lesssim 2$~TeV ($2000$~TeV)
for lepton number conserving (violating) operators. For comparison, let us add that searches for $h\to NN$ triggered by $\mathcal{O}_{NNH}$, with $N\to q\overline{q} \ell$ are expected to test scales as large as $\sim 100$ TeV~\cite{Caputo:2017pit}. Likewise, top decays into $b\ell N$, mediated by four-fermion operators and with $N$ long-lived, have been shown to probe only $\Lambda\lesssim 1$ TeV~\cite{Alcaide:2019pnf}.

\section*{Acknowledgements}
We thank Peter Galler and Jos\'e Santiago for helpful discussions. 
This research was supported by the Munich Institute for Astro- 
and Particle Physics (MIAPP) of the DFG Excellence Cluster Origins 
(\url{www.origins-cluster.de}).
CE acknowledges support by the UK Science and Technology Facilities Council 
(STFC), under grant ST/P000746/1.
MC is supported by the Spanish MINECO 
under the Juan de la Cierva programme 
and by the Royal Society under the Newton International Fellowship programme. 
AT is supported by the European Research Council 
under ERC Grant NuMass (FP7-IDEAS-ERC ERC-CG 617143).
AT and JB acknowledge funding
from the European Union's Horizon 2020 research and innovation programme 
under the Marie Sk\l{}odowska-Curie grant agreements 
No 674896 (ITN Elusives) and No 722104 (MCnetITN3), respectively.

\appendix
\section{Model}
\label{sec:app}
Let us consider the SM extended with two vector-like fermions $X_E\sim (\mathbf{1}, \mathbf{2})_{1/2}$, $X_N\sim (\mathbf{1}, \mathbf{1})_1$ and a singly-charged scalar $\varphi\sim (\mathbf{1},\mathbf{1})_1$. The numbers in parentheses and the subindex represent the quantum numbers under $(SU(3)_c, SU(2)_L)$ and the hypercharge, respectively. We also assume that these new fields are odd under a $\mathbb{Z}_2$ symmetry under which all SM fields as well as $N$ are even.

The new relevant Lagrangian reads
\begin{align}
L &= \overline{X_E}(i\slashed{D}-M_{X_E})X_E + \overline{X_N}(i\slashed{D}-M_{X_N})X_N \nonumber\\
&+(D_\mu \varphi)^\ast (D^\mu \varphi)-M_\varphi^2\varphi^\ast\varphi 
- \lambda_{\varphi H} (\varphi^\ast \varphi) (H^\dagger H) \nonumber\\
&+ \bigg[g_X \overline{X_E} \tilde{H} X_N + g_L \overline{X_E}\varphi L + g_N\overline{X_N}\varphi N + \text{h.c.}\bigg]~.
\end{align}
%

Let us focus on the regime $M_{X_E}, M_{X_N}, M_\varphi \sim M \gg v$, $g_N\ll g_L, g_X$. The new particles can be integrated out before EWSB by matching (off-shell) amplitudes in the UV to the corresponding amplitudes in the EFT. One can easily check that tree-level operators vanish.

\begin{figure*}[t]
\centering
\includegraphics[width=\textwidth]{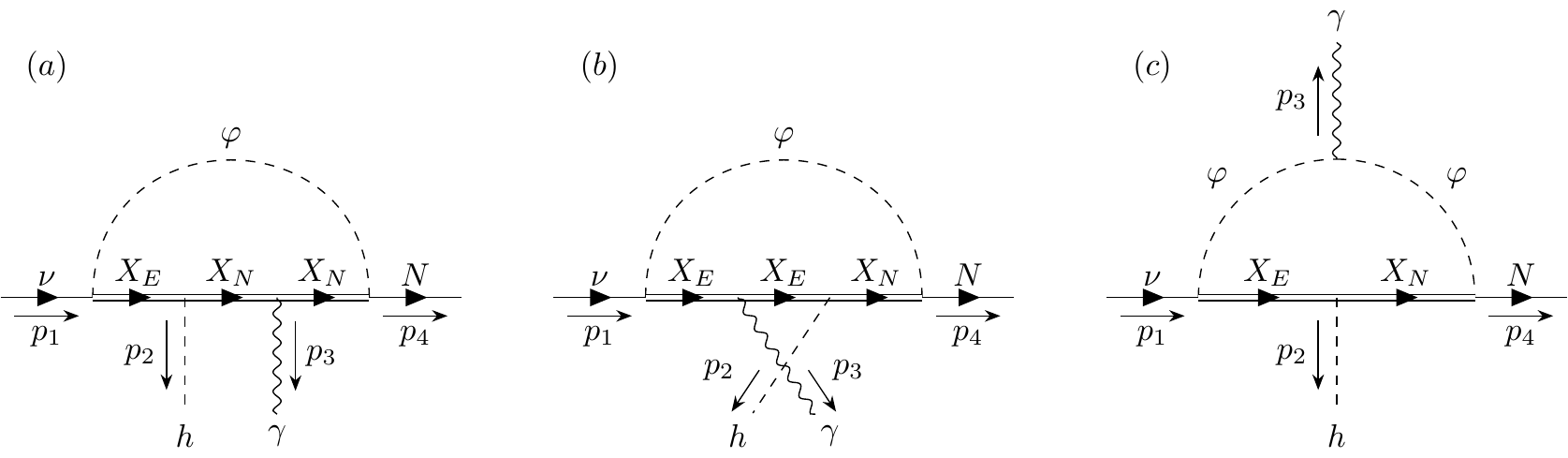}
\caption{\it Leading diagrams contributing to the amplitude with the Higgs, 
one photon, $\nu_L$ and $N$ before EWSB. 
The charge of the component of $X_E$, 
as well as the charges of $X_N$ and $\varphi$, are implicit.}
\label{fig:matching}
\end{figure*}

Therefore, we concentrate first on the amplitude 
given by the diagrams in Fig.~\ref{fig:matching}. 
%
Using $p_2, p_3$ and $p_4$ as independent four-momenta ($p_1 = p_2+p_3+p_4$), and to first order in $p_i$ we get:
\begin{widetext}
\begin{align}\nonumber
i\mathcal{M}_{a+b} = \frac{g_L g_X g_N e}{\sqrt{2}}~\overline{u}(p_4)P_L \bigg\lbrace 
&2 \left[2 B_4 - 3 C_5 + M^2 (A_4 - 10 B_5)\right]p_2^\mu 
+ 4 \left[B_4 - 3 C_5 - 4 M^2 B_5\right]p_3^\mu \\
&+6 \left[2 B_4 - 3 C_5 + M^2 (A_4 - 6 B_5) \right] p_4^\mu 
+ \left[2 B_4 - 3 C_5 + M^2 (A_4 +2 B_5)\right]\gamma^\mu \slashed{p}_2 \nonumber \\
&+ \left[2 B_4 + 3 C_5 + M^2 (3 A_4 - 2 B_5)\right]\gamma^\mu \slashed{p}_3
\bigg\rbrace u(p_1) \epsilon_\mu^*(p_3)\,,
\end{align}
%
and
\begin{align}
i\mathcal{M}_{c} = \frac{g_L g_X g_N e}{\sqrt{2}}~\overline{u}(p_4)P_L \bigg\lbrace 
&-2 \left[3 C_5+2 M^2 B_5\right] p_2^\mu 
  + \left[8 B_4-18 C_5+M^2(A_4 - 12 B_5)\right] p_3^\mu \\\nonumber
&+ 4\left[B_4-3 C_5-2 M^2 B_5\right] p_4^\mu 
  + 2 B_4 \gamma^\mu \slashed{p}_2 \bigg\rbrace u(p_1) \epsilon_\mu^*(p_3)\,.
\end{align}
%
$A_n, B_n$ and $C_n$ are four dimensional integrals defined by 
(see \textit{e.g.}~\cite{Peskin:1995ev})
\begin{align}
 \int \frac{d^4k}{(2\pi)^4} \frac{1}{(k^2 -M^2)^n} &=  A_n\,,\\
 \int \frac{d^4k}{(2\pi)^4} \frac{k_\mu k_\nu}{(k^2-M^2)^n} &=   B_n g_{\mu\nu}\,,\\
  \int \frac{d^4k}{(2\pi)^4} \frac{k_\mu k_\nu k_\rho k_\sigma}{(k^2-M^2)^n} &=  \frac{1}{4} C_n(g_{\mu\nu}g_{\rho\sigma} + g_{\mu\rho}g_{\nu\sigma} + g_{\mu\sigma}g_{\nu\rho})\,.
\end{align}
%
\end{widetext}
%
\begin{SCfigure*}[20][t]
\centering
\includegraphics[width=0.7\textwidth]{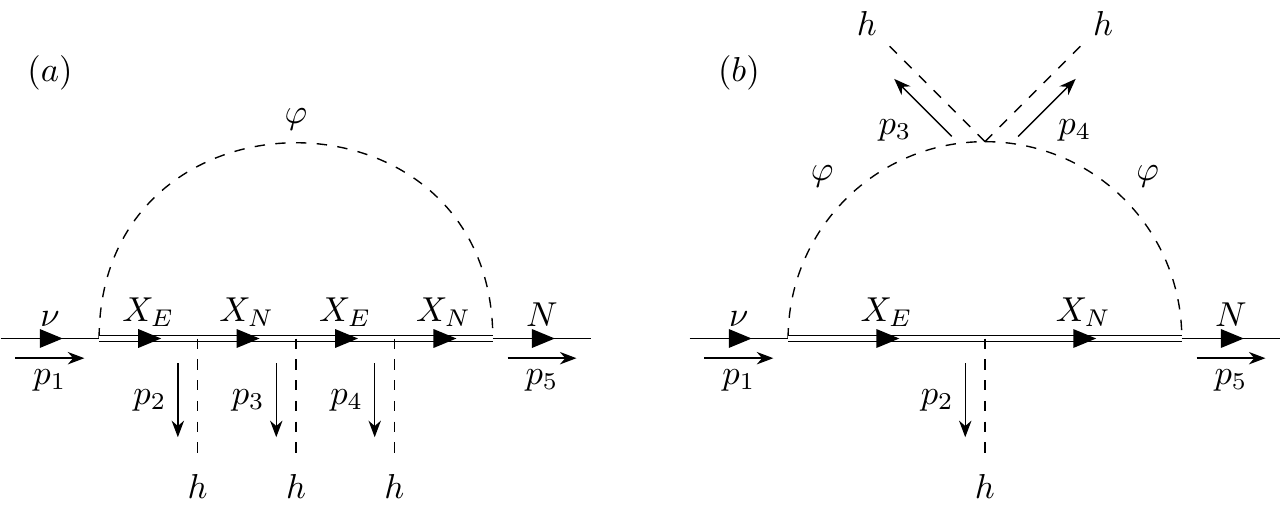}
\hfill
\caption{\it Leading diagrams contributing to the amplitude with three Higgses, 
$\nu_L$ and $N$ before EWSB. 
The charge of the component of $X_E$, 
as well as the charges of $X_N$ and $\varphi$, are implicit. 
Permutations of $p_2$, $p_3$ and $p_4$ 
have been taken into account in Eqs.~\eqref{eq:M'a} and \eqref{eq:M'b}.
}
\label{fig:matching2}
\end{SCfigure*}
%
Explicitly:
\begin{align}
 A_n &= \frac{(-1)^n i}{16 \pi^2 M^{2n-4}}\frac{\Gamma(n-2)}{\Gamma(n)}\,,\\
 B_n &= \frac{(-1)^{n-1} i}{32 \pi^2 M^{2n-6}}\frac{\Gamma(n-3)}{\Gamma(n)}\,,\\
 C_n &= \frac{3 (-1)^n i}{8 \pi^2 M^{2n-8}}\frac{\Gamma(n-4)}{\Gamma(n)}\,.
\end{align}
%
Adding all pieces together and simplifying further, we get:
\begin{align}
 i\mathcal{M}_{UV} &= i\mathcal{M}_{a+b}+i\mathcal{M}_c \nonumber\\
 &=  \frac{i g_L g_X g_N e}{96\sqrt{2}\pi^2 M^2}~\overline{u}(p_4)P_L 
 \left[\gamma^\mu\slashed{p}_3 - p_3^\mu\right] u(p_1) \epsilon_\mu^*(p_3)\,.
\end{align}
%
The only operator in the IR that contributes directly to the same amplitude is $\mathcal{O}_{NA}$; it reads:
\begin{equation}
 i\mathcal{M}_{IR} = \frac{i}{\Lambda^2} \sqrt{2} \alpha_{NA} \overline{u}(p_4)P_L \left[\gamma^\mu\slashed{p}_3 - p_3^\mu\right] u(p_1) \epsilon_\mu^*(p_3)\,.
\end{equation}
%
Upon requiring $\mathcal{M}_{UV} = \mathcal{M}_{IR}$ we finally obtain
\begin{equation}
 \frac{\alpha_{NA}}{\Lambda^2} = \frac{g_L g_X g_N e}{192\pi^2 M^2} ~.
\end{equation}
%

In order to obtain $\alpha_{LNH}$, we compute the amplitude given by 
the diagrams in Fig.~\ref{fig:matching2} to zero momentum. 
We have:
\begin{align}
 i\mathcal{M}_a^\prime &= -\frac{3}{\sqrt{2}} g_N g_X^3 g_L \overline{u}(p_5) P_L \big[6 C_5 + 24 M^2 B_5 \nonumber \\
 &\phantom{{}={}}+ M^4 A_5\big] u(p_1)\,, 
 \label{eq:M'a} \\
i\mathcal{M}_b^\prime &= - \frac{3}{\sqrt{2}} \lambda_{\varphi H} g_X g_L g_N 
\overline{u}(p_5) P_L \left[4 B_4 + M^2 A_4\right] u(p_1)\,.
\label{eq:M'b}
\end{align}
%
Adding both contributions, we obtain
\begin{align}
i\mathcal{M}_{UV}^\prime &= i\mathcal{M}^\prime_a+i\mathcal{M}^\prime_b \nonumber\\
&= \frac{ig_N g_X g_L}{32\sqrt{2}\pi^2M^2}
\left[\lambda_{\varphi H}-g_X^2\right]
\overline{u}(p_5) P_L u(p_1)\,.
\end{align}
%
In the EFT we obtain instead:
\begin{equation}
 i\mathcal{M}_{IR}^\prime = \frac{3i\alpha_{LNH}}{\sqrt{2} \Lambda^2} \overline{u}(p_5) P_L u(p_1)~,
\end{equation}
from where we obtain
\begin{equation}
\frac{\alpha_{LNH}}{\Lambda^2} = 
\frac{g_N g_L g_X}{96\pi^2M^2}
\left[\lambda_{\varphi H}-g_X^2\right].
\end{equation}
%

Redundant operators can be generated in the off-shell matching, 
therefore potentially contributing to $\mathcal{O}_{NA}$ and $\mathcal{O}_{LNH}$ 
after using the equations of motion of the SM$+N$. 
The relevant list of such operators reads:
\begin{align}
 \mathcal{O}_1 &= (\overline{L} N) D^2 \tilde{H}\,,\\
 \mathcal{O}_2 &= \overline{L} D_\mu N D^\mu \tilde{H}\,,\\
 \mathcal{O}_3 &= i\overline{L}\sigma^{\mu\nu} D_\mu N D_\nu \tilde{H}\,,\\
 \mathcal{O}_4 &= \overline{L} D^2 N \tilde{H}\,.
\end{align}
%
(The addition of hermitian conjugate is implied.) Other operators (not related 
to the ones above by algebraic identities or integration by parts) involve two 
copies of $N$, as we have cross checked using 
\texttt{BasisGen}~\cite{Criado:2019ugp}. Therefore, their Wilson coefficients 
are suppressed by two powers of $g_N$ and therefore negligible within our 
approximation $g_N\ll g_L, g_X$.

We expect contributions to $\mathcal{O}_{NA}$ to be small.
Likewise, further contributions to $\alpha_{LNH}$ come from the equations of motion of the Higgs~\cite{Grzadkowski:2010es}:
\begin{equation}
 D^\mu D_\mu H = \mu^2 H-\lambda_H (H^\dagger H) H-y_f \overline{f_L}f_R~.
\end{equation}
%
They are therefore suppressed by a further factor of $\lambda_H\sim 0.1$ and therefore negligible. In summary, assuming $\Lambda = M$, in good approximation:
\begin{equation}
 \alpha_{NA} \sim \frac{g_L g_X g_N e}{192\pi^2} ~, 
 \quad 
 \alpha_{LNH} \sim \frac{g_N g_L g_X}{96\pi^2}
\left[\lambda_{\varphi H}-g_X^2\right].
\end{equation}
%
We note that in the strongly coupled regime, and for $m_N\sim 1$ GeV 
and $\Lambda = 1$~TeV, $N$ decays 
effectively promptly  within the detector (see Fig.~\ref{fig:contur}), and 
$\alpha_{LNH}$ is within the reach of our analyses (see 
Tab.~\ref{tab:limits}). For example, neglecting $\lambda_{\varphi H}$ and for 
$g_L = g_X = \sqrt{4\pi}$ and $g_N = 1$, we get $\alpha_{NA} \sim 0.002$ and 
$\alpha_{LNH} \sim -0.17$. $\alpha_{NA}$ grows up to $\sim 0.03$ for $g_L=g_X = 
4\pi$.

\bibliography{references} 

\begin{thebibliography}{74}
\expandafter\ifx\csname natexlab\endcsname\relax\def\natexlab#1{#1}\fi
\expandafter\ifx\csname bibnamefont\endcsname\relax
  \def\bibnamefont#1{#1}\fi
\expandafter\ifx\csname bibfnamefont\endcsname\relax
  \def\bibfnamefont#1{#1}\fi
\expandafter\ifx\csname citenamefont\endcsname\relax
  \def\citenamefont#1{#1}\fi
\expandafter\ifx\csname url\endcsname\relax
  \def\url#1{\texttt{#1}}\fi
\expandafter\ifx\csname urlprefix\endcsname\relax\def\urlprefix{URL }\fi
\providecommand{\bibinfo}[2]{#2}
\providecommand{\eprint}[2][]{\url{#2}}

\bibitem[{\citenamefont{Grzadkowski et~al.}(2010)\citenamefont{Grzadkowski,
  Iskrzynski, Misiak, and Rosiek}}]{Grzadkowski:2010es}
\bibinfo{author}{\bibfnamefont{B.}~\bibnamefont{Grzadkowski}},
  \bibinfo{author}{\bibfnamefont{M.}~\bibnamefont{Iskrzynski}},
  \bibinfo{author}{\bibfnamefont{M.}~\bibnamefont{Misiak}}, \bibnamefont{and}
  \bibinfo{author}{\bibfnamefont{J.}~\bibnamefont{Rosiek}},
  \bibinfo{journal}{JHEP} \textbf{\bibinfo{volume}{10}}, \bibinfo{pages}{085}
  (\bibinfo{year}{2010}), \eprint{1008.4884}.

\bibitem[{\citenamefont{Brivio and Trott}(2019)}]{Brivio:2017vri}
\bibinfo{author}{\bibfnamefont{I.}~\bibnamefont{Brivio}} \bibnamefont{and}
  \bibinfo{author}{\bibfnamefont{M.}~\bibnamefont{Trott}},
  \bibinfo{journal}{Phys. Rept.} \textbf{\bibinfo{volume}{793}},
  \bibinfo{pages}{1} (\bibinfo{year}{2019}), \eprint{1706.08945}.

\bibitem[{\citenamefont{Egana-Ugrinovic
  et~al.}(2018)\citenamefont{Egana-Ugrinovic, Low, and
  Ruderman}}]{Egana-Ugrinovic:2018roi}
\bibinfo{author}{\bibfnamefont{D.}~\bibnamefont{Egana-Ugrinovic}},
  \bibinfo{author}{\bibfnamefont{M.}~\bibnamefont{Low}}, \bibnamefont{and}
  \bibinfo{author}{\bibfnamefont{J.~T.} \bibnamefont{Ruderman}},
  \bibinfo{journal}{JHEP} \textbf{\bibinfo{volume}{05}}, \bibinfo{pages}{012}
  (\bibinfo{year}{2018}), \eprint{1801.05432}.

\bibitem[{\citenamefont{Alcaide and Mileo}(2019)}]{Alcaide:2019kdr}
\bibinfo{author}{\bibfnamefont{J.}~\bibnamefont{Alcaide}} \bibnamefont{and}
  \bibinfo{author}{\bibfnamefont{N.~I.} \bibnamefont{Mileo}}
  (\bibinfo{year}{2019}), \eprint{1906.08685}.

\bibitem[{\citenamefont{del Aguila et~al.}(2009)\citenamefont{del Aguila,
  Bar-Shalom, Soni, and Wudka}}]{delAguila:2008ir}
\bibinfo{author}{\bibfnamefont{F.}~\bibnamefont{del Aguila}},
  \bibinfo{author}{\bibfnamefont{S.}~\bibnamefont{Bar-Shalom}},
  \bibinfo{author}{\bibfnamefont{A.}~\bibnamefont{Soni}}, \bibnamefont{and}
  \bibinfo{author}{\bibfnamefont{J.}~\bibnamefont{Wudka}},
  \bibinfo{journal}{Phys. Lett.} \textbf{\bibinfo{volume}{B670}},
  \bibinfo{pages}{399} (\bibinfo{year}{2009}), \eprint{0806.0876}.

\bibitem[{\citenamefont{Aparici et~al.}(2009)\citenamefont{Aparici, Kim,
  Santamaria, and Wudka}}]{Aparici:2009fh}
\bibinfo{author}{\bibfnamefont{A.}~\bibnamefont{Aparici}},
  \bibinfo{author}{\bibfnamefont{K.}~\bibnamefont{Kim}},
  \bibinfo{author}{\bibfnamefont{A.}~\bibnamefont{Santamaria}},
  \bibnamefont{and} \bibinfo{author}{\bibfnamefont{J.}~\bibnamefont{Wudka}},
  \bibinfo{journal}{Phys. Rev.} \textbf{\bibinfo{volume}{D80}},
  \bibinfo{pages}{013010} (\bibinfo{year}{2009}), \eprint{0904.3244}.

\bibitem[{\citenamefont{Bhattacharya and Wudka}(2016)}]{Bhattacharya:2015vja}
\bibinfo{author}{\bibfnamefont{S.}~\bibnamefont{Bhattacharya}}
  \bibnamefont{and} \bibinfo{author}{\bibfnamefont{J.}~\bibnamefont{Wudka}},
  \bibinfo{journal}{Phys. Rev.} \textbf{\bibinfo{volume}{D94}},
  \bibinfo{pages}{055022} (\bibinfo{year}{2016}), \bibinfo{note}{[Erratum:
  Phys. Rev.D95,no.3,039904(2017)]}, \eprint{1505.05264}.

\bibitem[{\citenamefont{Liao and Ma}(2017)}]{Liao:2016qyd}
\bibinfo{author}{\bibfnamefont{Y.}~\bibnamefont{Liao}} \bibnamefont{and}
  \bibinfo{author}{\bibfnamefont{X.-D.} \bibnamefont{Ma}},
  \bibinfo{journal}{Phys. Rev.} \textbf{\bibinfo{volume}{D96}},
  \bibinfo{pages}{015012} (\bibinfo{year}{2017}), \eprint{1612.04527}.

\bibitem[{\citenamefont{Franceschini et~al.}(2016)\citenamefont{Franceschini,
  Giudice, Kamenik, McCullough, Riva, Strumia, and
  Torre}}]{Franceschini:2016gxv}
\bibinfo{author}{\bibfnamefont{R.}~\bibnamefont{Franceschini}},
  \bibinfo{author}{\bibfnamefont{G.~F.} \bibnamefont{Giudice}},
  \bibinfo{author}{\bibfnamefont{J.~F.} \bibnamefont{Kamenik}},
  \bibinfo{author}{\bibfnamefont{M.}~\bibnamefont{McCullough}},
  \bibinfo{author}{\bibfnamefont{F.}~\bibnamefont{Riva}},
  \bibinfo{author}{\bibfnamefont{A.}~\bibnamefont{Strumia}}, \bibnamefont{and}
  \bibinfo{author}{\bibfnamefont{R.}~\bibnamefont{Torre}},
  \bibinfo{journal}{JHEP} \textbf{\bibinfo{volume}{07}}, \bibinfo{pages}{150}
  (\bibinfo{year}{2016}), \eprint{1604.06446}.

\bibitem[{\citenamefont{Gripaios and Sutherland}(2016)}]{Gripaios:2016xuo}
\bibinfo{author}{\bibfnamefont{B.}~\bibnamefont{Gripaios}} \bibnamefont{and}
  \bibinfo{author}{\bibfnamefont{D.}~\bibnamefont{Sutherland}},
  \bibinfo{journal}{JHEP} \textbf{\bibinfo{volume}{08}}, \bibinfo{pages}{103}
  (\bibinfo{year}{2016}), \eprint{1604.07365}.

\bibitem[{\citenamefont{Minkowski}(1977)}]{Minkowski:1977sc}
\bibinfo{author}{\bibfnamefont{P.}~\bibnamefont{Minkowski}},
  \bibinfo{journal}{Phys. Lett.} \textbf{\bibinfo{volume}{67B}},
  \bibinfo{pages}{421} (\bibinfo{year}{1977}).

\bibitem[{\citenamefont{Yanagida}(1979)}]{Yanagida:1979as}
\bibinfo{author}{\bibfnamefont{T.}~\bibnamefont{Yanagida}},
  \bibinfo{journal}{Conf. Proc.} \textbf{\bibinfo{volume}{C7902131}},
  \bibinfo{pages}{95} (\bibinfo{year}{1979}).

\bibitem[{\citenamefont{Gell-Mann et~al.}(1979)\citenamefont{Gell-Mann, Ramond,
  and Slansky}}]{GellMann:1980vs}
\bibinfo{author}{\bibfnamefont{M.}~\bibnamefont{Gell-Mann}},
  \bibinfo{author}{\bibfnamefont{P.}~\bibnamefont{Ramond}}, \bibnamefont{and}
  \bibinfo{author}{\bibfnamefont{R.}~\bibnamefont{Slansky}},
  \bibinfo{journal}{Conf. Proc.} \textbf{\bibinfo{volume}{C790927}},
  \bibinfo{pages}{315} (\bibinfo{year}{1979}), \eprint{1306.4669}.

\bibitem[{\citenamefont{Glashow}(1980)}]{Glashow:1979nm}
\bibinfo{author}{\bibfnamefont{S.~L.} \bibnamefont{Glashow}},
  \bibinfo{journal}{NATO Sci. Ser. B} \textbf{\bibinfo{volume}{61}},
  \bibinfo{pages}{687} (\bibinfo{year}{1980}).

\bibitem[{\citenamefont{Mohapatra and Senjanovic}(1980)}]{Mohapatra:1979ia}
\bibinfo{author}{\bibfnamefont{R.~N.} \bibnamefont{Mohapatra}}
  \bibnamefont{and}
  \bibinfo{author}{\bibfnamefont{G.}~\bibnamefont{Senjanovic}},
  \bibinfo{journal}{Phys. Rev. Lett.} \textbf{\bibinfo{volume}{44}},
  \bibinfo{pages}{912} (\bibinfo{year}{1980}).

\bibitem[{\citenamefont{Mohapatra}(1986)}]{Mohapatra:1986aw}
\bibinfo{author}{\bibfnamefont{R.~N.} \bibnamefont{Mohapatra}},
  \bibinfo{journal}{Phys. Rev. Lett.} \textbf{\bibinfo{volume}{56}},
  \bibinfo{pages}{561} (\bibinfo{year}{1986}).

\bibitem[{\citenamefont{Mohapatra and Valle}(1986)}]{Mohapatra:1986bd}
\bibinfo{author}{\bibfnamefont{R.~N.} \bibnamefont{Mohapatra}}
  \bibnamefont{and} \bibinfo{author}{\bibfnamefont{J.~W.~F.}
  \bibnamefont{Valle}}, \bibinfo{journal}{Phys. Rev.}
  \textbf{\bibinfo{volume}{D34}}, \bibinfo{pages}{1642} (\bibinfo{year}{1986}).

\bibitem[{\citenamefont{Bernabeu et~al.}(1987)\citenamefont{Bernabeu,
  Santamaria, Vidal, Mendez, and Valle}}]{Bernabeu:1987gr}
\bibinfo{author}{\bibfnamefont{J.}~\bibnamefont{Bernabeu}},
  \bibinfo{author}{\bibfnamefont{A.}~\bibnamefont{Santamaria}},
  \bibinfo{author}{\bibfnamefont{J.}~\bibnamefont{Vidal}},
  \bibinfo{author}{\bibfnamefont{A.}~\bibnamefont{Mendez}}, \bibnamefont{and}
  \bibinfo{author}{\bibfnamefont{J.~W.~F.} \bibnamefont{Valle}},
  \bibinfo{journal}{Phys. Lett.} \textbf{\bibinfo{volume}{B187}},
  \bibinfo{pages}{303} (\bibinfo{year}{1987}).

\bibitem[{\citenamefont{Akhmedov
  et~al.}(1996{\natexlab{a}})\citenamefont{Akhmedov, Lindner, Schnapka, and
  Valle}}]{Akhmedov:1995ip}
\bibinfo{author}{\bibfnamefont{E.~K.} \bibnamefont{Akhmedov}},
  \bibinfo{author}{\bibfnamefont{M.}~\bibnamefont{Lindner}},
  \bibinfo{author}{\bibfnamefont{E.}~\bibnamefont{Schnapka}}, \bibnamefont{and}
  \bibinfo{author}{\bibfnamefont{J.~W.~F.} \bibnamefont{Valle}},
  \bibinfo{journal}{Phys. Lett.} \textbf{\bibinfo{volume}{B368}},
  \bibinfo{pages}{270} (\bibinfo{year}{1996}{\natexlab{a}}),
  \eprint{hep-ph/9507275}.

\bibitem[{\citenamefont{Akhmedov
  et~al.}(1996{\natexlab{b}})\citenamefont{Akhmedov, Lindner, Schnapka, and
  Valle}}]{Akhmedov:1995vm}
\bibinfo{author}{\bibfnamefont{E.~K.} \bibnamefont{Akhmedov}},
  \bibinfo{author}{\bibfnamefont{M.}~\bibnamefont{Lindner}},
  \bibinfo{author}{\bibfnamefont{E.}~\bibnamefont{Schnapka}}, \bibnamefont{and}
  \bibinfo{author}{\bibfnamefont{J.~W.~F.} \bibnamefont{Valle}},
  \bibinfo{journal}{Phys. Rev.} \textbf{\bibinfo{volume}{D53}},
  \bibinfo{pages}{2752} (\bibinfo{year}{1996}{\natexlab{b}}),
  \eprint{hep-ph/9509255}.

\bibitem[{\citenamefont{Malinsky et~al.}(2005)\citenamefont{Malinsky, Romao,
  and Valle}}]{Malinsky:2005bi}
\bibinfo{author}{\bibfnamefont{M.}~\bibnamefont{Malinsky}},
  \bibinfo{author}{\bibfnamefont{J.~C.} \bibnamefont{Romao}}, \bibnamefont{and}
  \bibinfo{author}{\bibfnamefont{J.~W.~F.} \bibnamefont{Valle}},
  \bibinfo{journal}{Phys. Rev. Lett.} \textbf{\bibinfo{volume}{95}},
  \bibinfo{pages}{161801} (\bibinfo{year}{2005}), \eprint{hep-ph/0506296}.

\bibitem[{\citenamefont{Duarte et~al.}(2015)\citenamefont{Duarte, Peressutti,
  and Sampayo}}]{Duarte:2015iba}
\bibinfo{author}{\bibfnamefont{L.}~\bibnamefont{Duarte}},
  \bibinfo{author}{\bibfnamefont{J.}~\bibnamefont{Peressutti}},
  \bibnamefont{and} \bibinfo{author}{\bibfnamefont{O.~A.}
  \bibnamefont{Sampayo}}, \bibinfo{journal}{Phys. Rev.}
  \textbf{\bibinfo{volume}{D92}}, \bibinfo{pages}{093002}
  (\bibinfo{year}{2015}), \eprint{1508.01588}.

\bibitem[{\citenamefont{Duarte et~al.}(2018)\citenamefont{Duarte, Peressutti,
  and Sampayo}}]{Duarte:2016caz}
\bibinfo{author}{\bibfnamefont{L.}~\bibnamefont{Duarte}},
  \bibinfo{author}{\bibfnamefont{J.}~\bibnamefont{Peressutti}},
  \bibnamefont{and} \bibinfo{author}{\bibfnamefont{O.~A.}
  \bibnamefont{Sampayo}}, \bibinfo{journal}{J. Phys.}
  \textbf{\bibinfo{volume}{G45}}, \bibinfo{pages}{025001}
  (\bibinfo{year}{2018}), \eprint{1610.03894}.

\bibitem[{\citenamefont{Alcaide et~al.}(2019)\citenamefont{Alcaide, Banerjee,
  Chala, and Titov}}]{Alcaide:2019pnf}
\bibinfo{author}{\bibfnamefont{J.}~\bibnamefont{Alcaide}},
  \bibinfo{author}{\bibfnamefont{S.}~\bibnamefont{Banerjee}},
  \bibinfo{author}{\bibfnamefont{M.}~\bibnamefont{Chala}}, \bibnamefont{and}
  \bibinfo{author}{\bibfnamefont{A.}~\bibnamefont{Titov}},
  \bibinfo{journal}{JHEP} \textbf{\bibinfo{volume}{08}}, \bibinfo{pages}{031}
  (\bibinfo{year}{2019}), \eprint{1905.11375}.

\bibitem[{\citenamefont{Bar-Shalom et~al.}(2006)\citenamefont{Bar-Shalom,
  Deshpande, Eilam, Jiang, and Soni}}]{BarShalom:2006bv}
\bibinfo{author}{\bibfnamefont{S.}~\bibnamefont{Bar-Shalom}},
  \bibinfo{author}{\bibfnamefont{N.~G.} \bibnamefont{Deshpande}},
  \bibinfo{author}{\bibfnamefont{G.}~\bibnamefont{Eilam}},
  \bibinfo{author}{\bibfnamefont{J.}~\bibnamefont{Jiang}}, \bibnamefont{and}
  \bibinfo{author}{\bibfnamefont{A.}~\bibnamefont{Soni}},
  \bibinfo{journal}{Phys. Lett.} \textbf{\bibinfo{volume}{B643}},
  \bibinfo{pages}{342} (\bibinfo{year}{2006}), \eprint{hep-ph/0608309}.

\bibitem[{\citenamefont{Cveti{\v c} et~al.}(2019)\citenamefont{Cveti{\v c},
  Das, and Zamora-Sa{\'a}}}]{Cvetic:2018elt}
\bibinfo{author}{\bibfnamefont{G.}~\bibnamefont{Cveti{\v c}}},
  \bibinfo{author}{\bibfnamefont{A.}~\bibnamefont{Das}}, \bibnamefont{and}
  \bibinfo{author}{\bibfnamefont{J.}~\bibnamefont{Zamora-Sa{\'a}}},
  \bibinfo{journal}{J. Phys.} \textbf{\bibinfo{volume}{G46}},
  \bibinfo{pages}{075002} (\bibinfo{year}{2019}), \eprint{1805.00070}.

\bibitem[{\citenamefont{Caputo et~al.}(2017)\citenamefont{Caputo, Hernandez,
  Lopez-Pavon, and Salvado}}]{Caputo:2017pit}
\bibinfo{author}{\bibfnamefont{A.}~\bibnamefont{Caputo}},
  \bibinfo{author}{\bibfnamefont{P.}~\bibnamefont{Hernandez}},
  \bibinfo{author}{\bibfnamefont{J.}~\bibnamefont{Lopez-Pavon}},
  \bibnamefont{and} \bibinfo{author}{\bibfnamefont{J.}~\bibnamefont{Salvado}},
  \bibinfo{journal}{JHEP} \textbf{\bibinfo{volume}{06}}, \bibinfo{pages}{112}
  (\bibinfo{year}{2017}), \eprint{1704.08721}.

\bibitem[{\citenamefont{Accomando et~al.}(2017)\citenamefont{Accomando,
  Delle~Rose, Moretti, Olaiya, and
  Shepherd-Themistocleous}}]{Accomando:2016rpc}
\bibinfo{author}{\bibfnamefont{E.}~\bibnamefont{Accomando}},
  \bibinfo{author}{\bibfnamefont{L.}~\bibnamefont{Delle~Rose}},
  \bibinfo{author}{\bibfnamefont{S.}~\bibnamefont{Moretti}},
  \bibinfo{author}{\bibfnamefont{E.}~\bibnamefont{Olaiya}}, \bibnamefont{and}
  \bibinfo{author}{\bibfnamefont{C.~H.} \bibnamefont{Shepherd-Themistocleous}},
  \bibinfo{journal}{JHEP} \textbf{\bibinfo{volume}{04}}, \bibinfo{pages}{081}
  (\bibinfo{year}{2017}), \eprint{1612.05977}.

\bibitem[{\citenamefont{Aghanim et~al.}(2018)}]{Aghanim:2018eyx}
\bibinfo{author}{\bibfnamefont{N.}~\bibnamefont{Aghanim}} \bibnamefont{et~al.}
  (\bibinfo{collaboration}{Planck}) (\bibinfo{year}{2018}),
  \eprint{1807.06209}.

\bibitem[{\citenamefont{Abazajian and Heeck}(2019)}]{Abazajian:2019oqj}
\bibinfo{author}{\bibfnamefont{K.~N.} \bibnamefont{Abazajian}}
  \bibnamefont{and} \bibinfo{author}{\bibfnamefont{J.}~\bibnamefont{Heeck}},
  \bibinfo{journal}{Phys. Rev.} \textbf{\bibinfo{volume}{D100}},
  \bibinfo{pages}{075027} (\bibinfo{year}{2019}), \eprint{1908.03286}.

\bibitem[{\citenamefont{Canas et~al.}(2016)\citenamefont{Canas, Miranda,
  Parada, Tortola, and Valle}}]{Canas:2015yoa}
\bibinfo{author}{\bibfnamefont{B.~C.} \bibnamefont{Canas}},
  \bibinfo{author}{\bibfnamefont{O.~G.} \bibnamefont{Miranda}},
  \bibinfo{author}{\bibfnamefont{A.}~\bibnamefont{Parada}},
  \bibinfo{author}{\bibfnamefont{M.}~\bibnamefont{Tortola}}, \bibnamefont{and}
  \bibinfo{author}{\bibfnamefont{J.~W.~F.} \bibnamefont{Valle}},
  \bibinfo{journal}{Phys. Lett.} \textbf{\bibinfo{volume}{B753}},
  \bibinfo{pages}{191} (\bibinfo{year}{2016}), \bibinfo{note}{[Addendum: Phys.
  Lett.B757,568(2016)]}, \eprint{1510.01684}.

\bibitem[{\citenamefont{Miranda et~al.}(2019)\citenamefont{Miranda, Papoulias,
  T{\'o}rtola, and Valle}}]{Miranda:2019wdy}
\bibinfo{author}{\bibfnamefont{O.~G.} \bibnamefont{Miranda}},
  \bibinfo{author}{\bibfnamefont{D.~K.} \bibnamefont{Papoulias}},
  \bibinfo{author}{\bibfnamefont{M.}~\bibnamefont{T{\'o}rtola}},
  \bibnamefont{and} \bibinfo{author}{\bibfnamefont{J.~W.~F.}
  \bibnamefont{Valle}}, \bibinfo{journal}{JHEP} \textbf{\bibinfo{volume}{07}},
  \bibinfo{pages}{103} (\bibinfo{year}{2019}), \eprint{1905.03750}.

\bibitem[{\citenamefont{Auerbach et~al.}(2001)}]{Auerbach:2001wg}
\bibinfo{author}{\bibfnamefont{L.~B.} \bibnamefont{Auerbach}}
  \bibnamefont{et~al.} (\bibinfo{collaboration}{LSND}), \bibinfo{journal}{Phys.
  Rev.} \textbf{\bibinfo{volume}{D63}}, \bibinfo{pages}{112001}
  (\bibinfo{year}{2001}), \eprint{hep-ex/0101039}.

\bibitem[{\citenamefont{Beda et~al.}(2012)\citenamefont{Beda, Brudanin, Egorov,
  Medvedev, Pogosov, Shirchenko, and Starostin}}]{Beda:2012zz}
\bibinfo{author}{\bibfnamefont{A.~G.} \bibnamefont{Beda}},
  \bibinfo{author}{\bibfnamefont{V.~B.} \bibnamefont{Brudanin}},
  \bibinfo{author}{\bibfnamefont{V.~G.} \bibnamefont{Egorov}},
  \bibinfo{author}{\bibfnamefont{D.~V.} \bibnamefont{Medvedev}},
  \bibinfo{author}{\bibfnamefont{V.~S.} \bibnamefont{Pogosov}},
  \bibinfo{author}{\bibfnamefont{M.~V.} \bibnamefont{Shirchenko}},
  \bibnamefont{and} \bibinfo{author}{\bibfnamefont{A.~S.}
  \bibnamefont{Starostin}}, \bibinfo{journal}{Adv. High Energy Phys.}
  \textbf{\bibinfo{volume}{2012}}, \bibinfo{pages}{350150}
  (\bibinfo{year}{2012}).

\bibitem[{\citenamefont{Bell et~al.}(2006)\citenamefont{Bell, Gorchtein,
  Ramsey-Musolf, Vogel, and Wang}}]{Bell:2006wi}
\bibinfo{author}{\bibfnamefont{N.~F.} \bibnamefont{Bell}},
  \bibinfo{author}{\bibfnamefont{M.}~\bibnamefont{Gorchtein}},
  \bibinfo{author}{\bibfnamefont{M.~J.} \bibnamefont{Ramsey-Musolf}},
  \bibinfo{author}{\bibfnamefont{P.}~\bibnamefont{Vogel}}, \bibnamefont{and}
  \bibinfo{author}{\bibfnamefont{P.}~\bibnamefont{Wang}},
  \bibinfo{journal}{Phys. Lett.} \textbf{\bibinfo{volume}{B642}},
  \bibinfo{pages}{377} (\bibinfo{year}{2006}), \eprint{hep-ph/0606248}.

\bibitem[{\citenamefont{Jenkins et~al.}(2018)\citenamefont{Jenkins, Manohar,
  and Stoffer}}]{Jenkins:2017dyc}
\bibinfo{author}{\bibfnamefont{E.~E.} \bibnamefont{Jenkins}},
  \bibinfo{author}{\bibfnamefont{A.~V.} \bibnamefont{Manohar}},
  \bibnamefont{and} \bibinfo{author}{\bibfnamefont{P.}~\bibnamefont{Stoffer}},
  \bibinfo{journal}{JHEP} \textbf{\bibinfo{volume}{01}}, \bibinfo{pages}{084}
  (\bibinfo{year}{2018}), \eprint{1711.05270}.

\bibitem[{\citenamefont{Tanabashi et~al.}(2018)}]{Tanabashi:2018oca}
\bibinfo{author}{\bibfnamefont{M.}~\bibnamefont{Tanabashi}}
  \bibnamefont{et~al.} (\bibinfo{collaboration}{Particle Data Group}),
  \bibinfo{journal}{Phys. Rev.} \textbf{\bibinfo{volume}{D98}},
  \bibinfo{pages}{030001} (\bibinfo{year}{2018}).

\bibitem[{\citenamefont{Adriani et~al.}(1992)}]{Adriani:1992iu}
\bibinfo{author}{\bibfnamefont{O.}~\bibnamefont{Adriani}} \bibnamefont{et~al.}
  (\bibinfo{collaboration}{L3}), \bibinfo{journal}{Phys. Lett.}
  \textbf{\bibinfo{volume}{B297}}, \bibinfo{pages}{469} (\bibinfo{year}{1992}).

\bibitem[{\citenamefont{Akers et~al.}(1995)}]{Akers:1994vh}
\bibinfo{author}{\bibfnamefont{R.}~\bibnamefont{Akers}} \bibnamefont{et~al.}
  (\bibinfo{collaboration}{OPAL}), \bibinfo{journal}{Z. Phys.}
  \textbf{\bibinfo{volume}{C65}}, \bibinfo{pages}{47} (\bibinfo{year}{1995}).

\bibitem[{\citenamefont{Abreu et~al.}(1997)}]{Abreu:1996vd}
\bibinfo{author}{\bibfnamefont{P.}~\bibnamefont{Abreu}} \bibnamefont{et~al.}
  (\bibinfo{collaboration}{DELPHI}), \bibinfo{journal}{Z. Phys.}
  \textbf{\bibinfo{volume}{C74}}, \bibinfo{pages}{577} (\bibinfo{year}{1997}).

\bibitem[{\citenamefont{Acciarri et~al.}(1997)}]{Acciarri:1997im}
\bibinfo{author}{\bibfnamefont{M.}~\bibnamefont{Acciarri}} \bibnamefont{et~al.}
  (\bibinfo{collaboration}{L3}), \bibinfo{journal}{Phys. Lett.}
  \textbf{\bibinfo{volume}{B412}}, \bibinfo{pages}{201} (\bibinfo{year}{1997}).

\bibitem[{\citenamefont{Magill et~al.}(2018)\citenamefont{Magill, Plestid,
  Pospelov, and Tsai}}]{Magill:2018jla}
\bibinfo{author}{\bibfnamefont{G.}~\bibnamefont{Magill}},
  \bibinfo{author}{\bibfnamefont{R.}~\bibnamefont{Plestid}},
  \bibinfo{author}{\bibfnamefont{M.}~\bibnamefont{Pospelov}}, \bibnamefont{and}
  \bibinfo{author}{\bibfnamefont{Y.-D.} \bibnamefont{Tsai}},
  \bibinfo{journal}{Phys. Rev.} \textbf{\bibinfo{volume}{D98}},
  \bibinfo{pages}{115015} (\bibinfo{year}{2018}), \eprint{1803.03262}.

\bibitem[{\citenamefont{Sirunyan
  et~al.}(2019{\natexlab{a}})}]{Sirunyan:2018dsf}
\bibinfo{author}{\bibfnamefont{A.~M.} \bibnamefont{Sirunyan}}
  \bibnamefont{et~al.} (\bibinfo{collaboration}{CMS}), \bibinfo{journal}{JHEP}
  \textbf{\bibinfo{volume}{02}}, \bibinfo{pages}{074}
  (\bibinfo{year}{2019}{\natexlab{a}}), \eprint{1810.00196}.

\bibitem[{\citenamefont{Brun and Rademakers}(1997)}]{Brun:1997pa}
\bibinfo{author}{\bibfnamefont{R.}~\bibnamefont{Brun}} \bibnamefont{and}
  \bibinfo{author}{\bibfnamefont{F.}~\bibnamefont{Rademakers}},
  \bibinfo{journal}{Nucl. Instrum. Meth.} \textbf{\bibinfo{volume}{A389}},
  \bibinfo{pages}{81} (\bibinfo{year}{1997}).

\bibitem[{\citenamefont{Antcheva et~al.}(2009)}]{Antcheva:2009zz}
\bibinfo{author}{\bibfnamefont{I.}~\bibnamefont{Antcheva}}
  \bibnamefont{et~al.}, \bibinfo{journal}{Comput. Phys. Commun.}
  \textbf{\bibinfo{volume}{180}}, \bibinfo{pages}{2499} (\bibinfo{year}{2009}),
  \eprint{1508.07749}.

\bibitem[{\citenamefont{Dobbs and Hansen}(2001)}]{Dobbs:2001ck}
\bibinfo{author}{\bibfnamefont{M.}~\bibnamefont{Dobbs}} \bibnamefont{and}
  \bibinfo{author}{\bibfnamefont{J.~B.} \bibnamefont{Hansen}},
  \bibinfo{journal}{Comput. Phys. Commun.} \textbf{\bibinfo{volume}{134}},
  \bibinfo{pages}{41} (\bibinfo{year}{2001}).

\bibitem[{\citenamefont{Cacciari et~al.}(2012)\citenamefont{Cacciari, Salam,
  and Soyez}}]{Cacciari:2011ma}
\bibinfo{author}{\bibfnamefont{M.}~\bibnamefont{Cacciari}},
  \bibinfo{author}{\bibfnamefont{G.~P.} \bibnamefont{Salam}}, \bibnamefont{and}
  \bibinfo{author}{\bibfnamefont{G.}~\bibnamefont{Soyez}},
  \bibinfo{journal}{Eur. Phys. J.} \textbf{\bibinfo{volume}{C72}},
  \bibinfo{pages}{1896} (\bibinfo{year}{2012}), \eprint{1111.6097}.

\bibitem[{\citenamefont{Hoya}(2018)}]{Hoya:2628323}
\bibinfo{author}{\bibfnamefont{J.}~\bibnamefont{Hoya}}
  (\bibinfo{collaboration}{ATLAS Collaboration}) (\bibinfo{year}{2018}),
  \urlprefix\url{https://cds.cern.ch/record/2628323}.

\bibitem[{\citenamefont{Read}(2000)}]{Read:2000ru}
\bibinfo{author}{\bibfnamefont{A.~L.} \bibnamefont{Read}}, in
  \emph{\bibinfo{booktitle}{{Workshop on confidence limits, CERN, Geneva,
  Switzerland, 17-18 Jan 2000: Proceedings}}} (\bibinfo{year}{2000}), pp.
  \bibinfo{pages}{81--101},
  \urlprefix\url{http://weblib.cern.ch/abstract?CERN-OPEN-2000-205}.

\bibitem[{\citenamefont{Read}(2002)}]{Read:2002hq}
\bibinfo{author}{\bibfnamefont{A.~L.} \bibnamefont{Read}}, \bibinfo{journal}{J.
  Phys.} \textbf{\bibinfo{volume}{G28}}, \bibinfo{pages}{2693}
  (\bibinfo{year}{2002}).

\bibitem[{\citenamefont{Butterworth et~al.}(2017)\citenamefont{Butterworth,
  Grellscheid, Kr{\"a}mer, Sarrazin, and Yallup}}]{Butterworth:2016sqg}
\bibinfo{author}{\bibfnamefont{J.~M.} \bibnamefont{Butterworth}},
  \bibinfo{author}{\bibfnamefont{D.}~\bibnamefont{Grellscheid}},
  \bibinfo{author}{\bibfnamefont{M.}~\bibnamefont{Kr{\"a}mer}},
  \bibinfo{author}{\bibfnamefont{B.}~\bibnamefont{Sarrazin}}, \bibnamefont{and}
  \bibinfo{author}{\bibfnamefont{D.}~\bibnamefont{Yallup}},
  \bibinfo{journal}{JHEP} \textbf{\bibinfo{volume}{03}}, \bibinfo{pages}{078}
  (\bibinfo{year}{2017}), \eprint{1606.05296}.

\bibitem[{\citenamefont{Bellm et~al.}(2016)}]{Bellm:2015jjp}
\bibinfo{author}{\bibfnamefont{J.}~\bibnamefont{Bellm}} \bibnamefont{et~al.},
  \bibinfo{journal}{Eur. Phys. J.} \textbf{\bibinfo{volume}{C76}},
  \bibinfo{pages}{196} (\bibinfo{year}{2016}), \eprint{1512.01178}.

\bibitem[{\citenamefont{Bahr et~al.}(2008)}]{Bahr:2008pv}
\bibinfo{author}{\bibfnamefont{M.}~\bibnamefont{Bahr}} \bibnamefont{et~al.},
  \bibinfo{journal}{Eur. Phys. J.} \textbf{\bibinfo{volume}{C58}},
  \bibinfo{pages}{639} (\bibinfo{year}{2008}), \eprint{0803.0883}.

\bibitem[{\citenamefont{Degrande et~al.}(2012)\citenamefont{Degrande, Duhr,
  Fuks, Grellscheid, Mattelaer, and Reiter}}]{Degrande:2011ua}
\bibinfo{author}{\bibfnamefont{C.}~\bibnamefont{Degrande}},
  \bibinfo{author}{\bibfnamefont{C.}~\bibnamefont{Duhr}},
  \bibinfo{author}{\bibfnamefont{B.}~\bibnamefont{Fuks}},
  \bibinfo{author}{\bibfnamefont{D.}~\bibnamefont{Grellscheid}},
  \bibinfo{author}{\bibfnamefont{O.}~\bibnamefont{Mattelaer}},
  \bibnamefont{and} \bibinfo{author}{\bibfnamefont{T.}~\bibnamefont{Reiter}},
  \bibinfo{journal}{Comput. Phys. Commun.} \textbf{\bibinfo{volume}{183}},
  \bibinfo{pages}{1201} (\bibinfo{year}{2012}), \eprint{1108.2040}.

\bibitem[{\citenamefont{Buckley et~al.}(2013)\citenamefont{Buckley,
  Butterworth, Lonnblad, Grellscheid, Hoeth, Monk, Schulz, and
  Siegert}}]{Buckley:2010ar}
\bibinfo{author}{\bibfnamefont{A.}~\bibnamefont{Buckley}},
  \bibinfo{author}{\bibfnamefont{J.}~\bibnamefont{Butterworth}},
  \bibinfo{author}{\bibfnamefont{L.}~\bibnamefont{Lonnblad}},
  \bibinfo{author}{\bibfnamefont{D.}~\bibnamefont{Grellscheid}},
  \bibinfo{author}{\bibfnamefont{H.}~\bibnamefont{Hoeth}},
  \bibinfo{author}{\bibfnamefont{J.}~\bibnamefont{Monk}},
  \bibinfo{author}{\bibfnamefont{H.}~\bibnamefont{Schulz}}, \bibnamefont{and}
  \bibinfo{author}{\bibfnamefont{F.}~\bibnamefont{Siegert}},
  \bibinfo{journal}{Comput. Phys. Commun.} \textbf{\bibinfo{volume}{184}},
  \bibinfo{pages}{2803} (\bibinfo{year}{2013}), \eprint{1003.0694}.

\bibitem[{\citenamefont{Maguire et~al.}(2017)\citenamefont{Maguire, Heinrich,
  and Watt}}]{Maguire:2017ypu}
\bibinfo{author}{\bibfnamefont{E.}~\bibnamefont{Maguire}},
  \bibinfo{author}{\bibfnamefont{L.}~\bibnamefont{Heinrich}}, \bibnamefont{and}
  \bibinfo{author}{\bibfnamefont{G.}~\bibnamefont{Watt}}, \bibinfo{journal}{J.
  Phys. Conf. Ser.} \textbf{\bibinfo{volume}{898}}, \bibinfo{pages}{102006}
  (\bibinfo{year}{2017}), \eprint{1704.05473}.

\bibitem[{\citenamefont{Aad et~al.}(2015{\natexlab{a}})}]{Aad:2014qxa}
\bibinfo{author}{\bibfnamefont{G.}~\bibnamefont{Aad}} \bibnamefont{et~al.}
  (\bibinfo{collaboration}{ATLAS}), \bibinfo{journal}{Eur. Phys. J.}
  \textbf{\bibinfo{volume}{C75}}, \bibinfo{pages}{82}
  (\bibinfo{year}{2015}{\natexlab{a}}), \eprint{1409.8639}.

\bibitem[{\citenamefont{Aad et~al.}(2013)}]{Aad:2013izg}
\bibinfo{author}{\bibfnamefont{G.}~\bibnamefont{Aad}} \bibnamefont{et~al.}
  (\bibinfo{collaboration}{ATLAS}), \bibinfo{journal}{Phys. Rev.}
  \textbf{\bibinfo{volume}{D87}}, \bibinfo{pages}{112003}
  (\bibinfo{year}{2013}), \bibinfo{note}{[Erratum: Phys.
  Rev.D91,no.11,119901(2015)]}, \eprint{1302.1283}.

\bibitem[{\citenamefont{Aad et~al.}(2016{\natexlab{a}})}]{Aad:2016xcr}
\bibinfo{author}{\bibfnamefont{G.}~\bibnamefont{Aad}} \bibnamefont{et~al.}
  (\bibinfo{collaboration}{ATLAS}), \bibinfo{journal}{JHEP}
  \textbf{\bibinfo{volume}{08}}, \bibinfo{pages}{005}
  (\bibinfo{year}{2016}{\natexlab{a}}), \eprint{1605.03495}.

\bibitem[{\citenamefont{Aaboud et~al.}(2018)}]{Aaboud:2017kff}
\bibinfo{author}{\bibfnamefont{M.}~\bibnamefont{Aaboud}} \bibnamefont{et~al.}
  (\bibinfo{collaboration}{ATLAS}), \bibinfo{journal}{Phys. Lett.}
  \textbf{\bibinfo{volume}{B780}}, \bibinfo{pages}{578} (\bibinfo{year}{2018}),
  \eprint{1801.00112}.

\bibitem[{\citenamefont{Aad et~al.}(2016{\natexlab{b}})}]{Aad:2016sau}
\bibinfo{author}{\bibfnamefont{G.}~\bibnamefont{Aad}} \bibnamefont{et~al.}
  (\bibinfo{collaboration}{ATLAS}), \bibinfo{journal}{Phys. Rev.}
  \textbf{\bibinfo{volume}{D93}}, \bibinfo{pages}{112002}
  (\bibinfo{year}{2016}{\natexlab{b}}), \eprint{1604.05232}.

\bibitem[{\citenamefont{Sirunyan
  et~al.}(2019{\natexlab{b}})}]{Sirunyan:2019xst}
\bibinfo{author}{\bibfnamefont{A.~M.} \bibnamefont{Sirunyan}}
  \bibnamefont{et~al.} (\bibinfo{collaboration}{CMS}), \bibinfo{journal}{JHEP}
  \textbf{\bibinfo{volume}{10}}, \bibinfo{pages}{139}
  (\bibinfo{year}{2019}{\natexlab{b}}), \eprint{1908.02699}.

\bibitem[{\citenamefont{Aad et~al.}(2012)}]{Aad:2012tfa}
\bibinfo{author}{\bibfnamefont{G.}~\bibnamefont{Aad}} \bibnamefont{et~al.}
  (\bibinfo{collaboration}{ATLAS}), \bibinfo{journal}{Phys. Lett.}
  \textbf{\bibinfo{volume}{B716}}, \bibinfo{pages}{1} (\bibinfo{year}{2012}),
  \eprint{1207.7214}.

\bibitem[{\citenamefont{Chatrchyan et~al.}(2012)}]{Chatrchyan:2012xdj}
\bibinfo{author}{\bibfnamefont{S.}~\bibnamefont{Chatrchyan}}
  \bibnamefont{et~al.} (\bibinfo{collaboration}{CMS}), \bibinfo{journal}{Phys.
  Lett.} \textbf{\bibinfo{volume}{B716}}, \bibinfo{pages}{30}
  (\bibinfo{year}{2012}), \eprint{1207.7235}.

\bibitem[{\citenamefont{Dittmaier et~al.}(2011)}]{Dittmaier:2011ti}
\bibinfo{author}{\bibfnamefont{S.}~\bibnamefont{Dittmaier}}
  \bibnamefont{et~al.} (\bibinfo{collaboration}{LHC Higgs Cross Section Working
  Group}) (\bibinfo{year}{2011}), \eprint{1101.0593}.

\bibitem[{\citenamefont{Dittmaier et~al.}(2012)}]{Dittmaier:2012vm}
\bibinfo{author}{\bibfnamefont{S.}~\bibnamefont{Dittmaier}}
  \bibnamefont{et~al.} (\bibinfo{year}{2012}), \eprint{1201.3084}.

\bibitem[{\citenamefont{Andersen et~al.}(2013)}]{Heinemeyer:2013tqa}
\bibinfo{author}{\bibfnamefont{J.~R.} \bibnamefont{Andersen}}
  \bibnamefont{et~al.} (\bibinfo{collaboration}{LHC Higgs Cross Section Working
  Group}) (\bibinfo{year}{2013}), \eprint{1307.1347}.

\bibitem[{\citenamefont{de~Florian et~al.}(2016)}]{deFlorian:2016spz}
\bibinfo{author}{\bibfnamefont{D.}~\bibnamefont{de~Florian}}
  \bibnamefont{et~al.} (\bibinfo{collaboration}{LHC Higgs Cross Section Working
  Group}) (\bibinfo{year}{2016}), \eprint{1610.07922}.

\bibitem[{\citenamefont{Aad et~al.}(2015{\natexlab{b}})}]{Aad:2015zhl}
\bibinfo{author}{\bibfnamefont{G.}~\bibnamefont{Aad}} \bibnamefont{et~al.}
  (\bibinfo{collaboration}{ATLAS, CMS}), \bibinfo{journal}{Phys. Rev. Lett.}
  \textbf{\bibinfo{volume}{114}}, \bibinfo{pages}{191803}
  (\bibinfo{year}{2015}{\natexlab{b}}), \eprint{1503.07589}.

\bibitem[{\citenamefont{Moneta et~al.}(2010)\citenamefont{Moneta, Belasco,
  Cranmer, Kreiss, Lazzaro, Piparo, Schott, Verkerke, and
  Wolf}}]{Moneta:2010pm}
\bibinfo{author}{\bibfnamefont{L.}~\bibnamefont{Moneta}},
  \bibinfo{author}{\bibfnamefont{K.}~\bibnamefont{Belasco}},
  \bibinfo{author}{\bibfnamefont{K.~S.} \bibnamefont{Cranmer}},
  \bibinfo{author}{\bibfnamefont{S.}~\bibnamefont{Kreiss}},
  \bibinfo{author}{\bibfnamefont{A.}~\bibnamefont{Lazzaro}},
  \bibinfo{author}{\bibfnamefont{D.}~\bibnamefont{Piparo}},
  \bibinfo{author}{\bibfnamefont{G.}~\bibnamefont{Schott}},
  \bibinfo{author}{\bibfnamefont{W.}~\bibnamefont{Verkerke}}, \bibnamefont{and}
  \bibinfo{author}{\bibfnamefont{M.}~\bibnamefont{Wolf}},
  \bibinfo{journal}{PoS} \textbf{\bibinfo{volume}{ACAT2010}},
  \bibinfo{pages}{057} (\bibinfo{year}{2010}), \eprint{1009.1003}.

\bibitem[{\citenamefont{Alwall et~al.}(2014)\citenamefont{Alwall, Frederix,
  Frixione, Hirschi, Maltoni, Mattelaer, Shao, Stelzer, Torrielli, and
  Zaro}}]{Alwall:2014hca}
\bibinfo{author}{\bibfnamefont{J.}~\bibnamefont{Alwall}},
  \bibinfo{author}{\bibfnamefont{R.}~\bibnamefont{Frederix}},
  \bibinfo{author}{\bibfnamefont{S.}~\bibnamefont{Frixione}},
  \bibinfo{author}{\bibfnamefont{V.}~\bibnamefont{Hirschi}},
  \bibinfo{author}{\bibfnamefont{F.}~\bibnamefont{Maltoni}},
  \bibinfo{author}{\bibfnamefont{O.}~\bibnamefont{Mattelaer}},
  \bibinfo{author}{\bibfnamefont{H.~S.} \bibnamefont{Shao}},
  \bibinfo{author}{\bibfnamefont{T.}~\bibnamefont{Stelzer}},
  \bibinfo{author}{\bibfnamefont{P.}~\bibnamefont{Torrielli}},
  \bibnamefont{and} \bibinfo{author}{\bibfnamefont{M.}~\bibnamefont{Zaro}},
  \bibinfo{journal}{JHEP} \textbf{\bibinfo{volume}{07}}, \bibinfo{pages}{079}
  (\bibinfo{year}{2014}), \eprint{1405.0301}.

\bibitem[{\citenamefont{Shrock and Suzuki}(1982)}]{Shrock:1982kd}
\bibinfo{author}{\bibfnamefont{R.~E.} \bibnamefont{Shrock}} \bibnamefont{and}
  \bibinfo{author}{\bibfnamefont{M.}~\bibnamefont{Suzuki}},
  \bibinfo{journal}{Phys. Lett.} \textbf{\bibinfo{volume}{110B}},
  \bibinfo{pages}{250} (\bibinfo{year}{1982}).

\bibitem[{\citenamefont{Peskin and Schroeder}(1995)}]{Peskin:1995ev}
\bibinfo{author}{\bibfnamefont{M.~E.} \bibnamefont{Peskin}} \bibnamefont{and}
  \bibinfo{author}{\bibfnamefont{D.~V.} \bibnamefont{Schroeder}},
  \emph{\bibinfo{title}{{An Introduction to quantum field theory}}}
  (\bibinfo{publisher}{Addison-Wesley}, \bibinfo{address}{Reading, USA},
  \bibinfo{year}{1995}), ISBN \bibinfo{isbn}{9780201503975, 0201503972},
  \urlprefix\url{http://www.slac.stanford.edu/~mpeskin/QFT.html}.

\bibitem[{\citenamefont{Criado}(2019)}]{Criado:2019ugp}
\bibinfo{author}{\bibfnamefont{J.~C.} \bibnamefont{Criado}},
  \bibinfo{journal}{Eur. Phys. J.} \textbf{\bibinfo{volume}{C79}},
  \bibinfo{pages}{256} (\bibinfo{year}{2019}), \eprint{1901.03501}.

\end{thebibliography}

\end{document}